\documentclass{article}

\RequirePackage{amsthm,amsmath,amsfonts,amssymb}
\RequirePackage[numbers]{natbib}
\RequirePackage{graphicx}
\usepackage{bm}
\usepackage{float}
\usepackage{multirow}
\usepackage{caption}
\usepackage{color}
\usepackage{amsmath,amssymb,amstext,amsthm,amsfonts}
\usepackage{dsfont}

\usepackage[pdftex,pagebackref=false]{hyperref}
\definecolor{background-color}{gray}{0.98}
\definecolor{steelblue}{rgb}{0.27, 0.51, 0.71}
\definecolor{brickred}{rgb}{0.8, 0.25, 0.33}
\definecolor{bluegray}{rgb}{0.4, 0.6, 0.8}
\definecolor{amethyst}{rgb}{0.6, 0.4, 0.8}

\hypersetup{
	plainpages=false,       
	unicode=false,          
	pdftoolbar=true,        
	pdfmenubar=true,        
	pdffitwindow=false,     
	pdfstartview={FitH},    
	pdftitle={A structural genomic model}, 
	pdfauthor={Chris Salahub},    
	pdfnewwindow=true,      
	colorlinks=true,        
	linkcolor=steelblue,         
	citecolor=brickred,        
	filecolor=magenta,      
	urlcolor=cyan           
}

\newcommand{\code}[1]{\texttt{#1}}

\newcommand*{\R}{\textsf{R}$~$}

\newcommand{\ve}[1]{\mathbf{#1}}           
\newcommand{\sv}[1]{\boldsymbol{#1}}   
\newcommand{\m}[1]{\mathbf{#1}}               
\newcommand{\sm}[1]{\boldsymbol{#1}}   
\newcommand{\tr}[1]{{#1}^{\mkern-1.5mu\mathsf{T}}}              

\newcommand{\ind}[2]{I_{#2} \left( #1 \right)}

\newcommand{\field}[1]{\mathbb{#1}}
\newcommand{\Reals}{\field{R}}

\newcommand{\Naturals}{\field{N}}

\title{A structural model of genome-wide association studies.}
\author{Christopher Salahub \\
  \textit{University of Waterloo} \\
  \href{mailto:csalahub@uwaterloo.ca}{csalahub@uwaterloo.ca}}

\begin{document}
	
\maketitle

\begin{abstract}
A structural genetic model incorporating a modern understanding of the genome and common practice in genome-wide association studies is derived mathematically. The model shows the Haldane map distance as a direct consequence of the structure of the genome. An expression for genetic correlation is derived under the model and compared to data resulting from the BSB mouse cross. A correlation test plot is introduced for this comparison and shows the close agreement of the model and empirical results. Noteworthy departures in this plot indicate regions which warrant further investigation.
\end{abstract}

\begin{center}
\begin{minipage}{0.8\textwidth}
\begin{center}
  \textbf{Keywords}: \textit{genetics, genomics, association, correlation, map distance, genome-wide association studies}
\end{center}
\end{minipage}
\end{center}

\section{Introduction} \label{sec:intro}

\begin{figure}[h]
  \begin{center}
  \includegraphics[scale = 1]{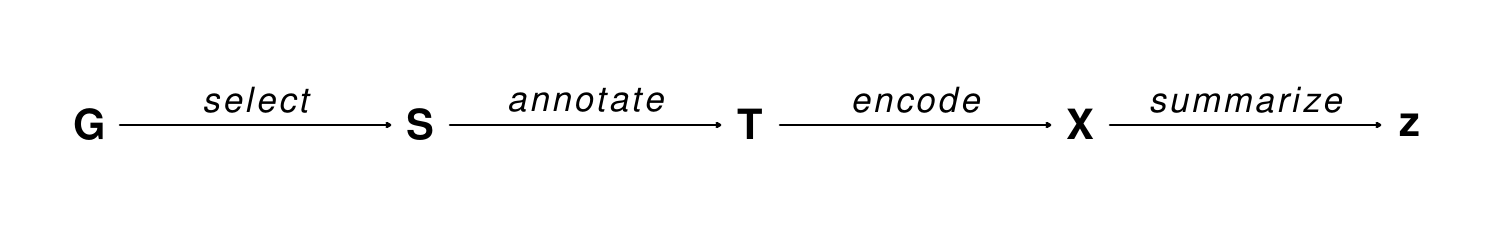}
  \caption{A structural model of genomic association
    studies.} \label{fig:modelDiagram}
  \end{center}
\end{figure}

Genetic research today routinely considers the entirety of a \emph{genome}, that is all heritable material potentially passed to offspring \cite{doergeetal1997search}, to identify regions strongly related to measured traits. The goal is to associate measured genome sequences, the \emph{genotype}, with physical characteristics, the \emph{phenotype}. Computational and methodological advances in the pursuit of these \emph{quantitative trait loci} (QTLs) have distinguished \emph{genomics} as its own field. Central to genomics is the \emph{genome-wide association study} (GWAS), where many \emph{markers}, sequences of nucleotides at known positions on the genome, are measured. Extracting useful results from markers and measured traits is a complicated task which has motivated decades of statistical and biological research. Interested individuals must sort through this voluminous research in order to understand genomics.

To aid in this, Figure \ref{fig:modelDiagram}
draws on the literature to present the structural model underlying the conversion of raw marker measurements into data to identify QTLs. The key steps of \emph{selection}, \emph{annotation}, \emph{encoding}, and \emph{summarization} are identified as maps between increasingly abstract representations of the genome. By highlighting these abstractions in plain language the model facilitates faster understanding of GWAS. While it is no replacement surveys such as \cite{uffelmannetal2021gwas, tametal2019benefits} or the literature they outline, it supplies a guiding structural framework of the field with exceptional explanatory power.

The model starts with $\m{G}$, the whole genome of an individual organism. Genetic information is stored in DNA, a long molecule consisting of a sequence of four \emph{nucleotide bases}: guanine, cytosine, adenine, and thymine. A \emph{diplodic} individual inherits one version or \emph{variant} of a complete DNA sequence from each parent, and so has two copies in all \emph{somatic} (i.e. non-reproductive) cells. Though it can be represented as one long sequence, DNA is actually organized into \emph{chromosomes}, separate strands of DNA which contain only a part of the sequence. As most genetic research concerns diplodic species, this is implicitly assumed.

It is usually not feasible or desirable to design a study around the measurement of all of $\m{G}$, and so the \emph{select} step chooses regions to measure. These regions are represented by $\m{S}$. Often $\m{S}$ consists of a series of \emph{single nucleotide polymorphisms} (SNPs), single nucleotide substitutions in a known sequence at a known position. In human studies this is supported by SNP databases such as dbSNP \cite{NCBIdbSNP} which document hundreds of millions of common SNPs in the human genome, of which perhaps 15 million occur frequently enough in humans to be useful \cite{koboldtetal2013next}. SNPs are sequenced by means of arrays capable of identifying roughly one million of these simultaneously \cite{laframboise2009, tametal2019benefits}. Selection is therefore necessary, motivated by previous literature and \emph{linkage disequilibrium}, the correlation between markers and other regions of the genome. Linkage disequilibrium also facilitates inference to regions outside of those selected in $\m{S}$. While third generation genome sequencing technologies allow for entire genomes to be sequenced, their persistent high costs and more than a decade of SNP array development leave arrays as the dominant measurement method \cite{heatherchain2016sequencers, hasinetal2017multi, uffelmannetal2021gwas}.

After selecting SNPs to obtain $\m{S}$, the data must be \emph{annotated}. The raw signal produced by a SNP array is fluorescence, with different degrees of fluorescence corresponding to different genotypes. Converting the fluorescent areas of an array to a genotype is a challenging problem and has developed in tandem with the arrays themselves. Early models used non-parametric clustering techniques on the signal from several array sections, but more complex hidden Markov and Bayesian models have also been developed \cite{laframboise2009}. Whatever method is used, the selected regions are assigned genotypes in $\m{T}$. Often these are denoted with capital or lowercase letters at each SNP, as in \cite{siegmundyakir2007, visschergoddard2019}.

Finally, relationships between $\m{T}$ and an observed trait or within $\m{T}$ itself are quantified by converting each annotated SNP to a number. GWAS first \emph{encode} each SNP variant with a numeric value and then \emph{summarize} both variants at each location into a single value. Typically no distinction is made between these steps: the dominance and additive summaries often move directly from an annotated genotype to a numeric value \cite{LanderBotstein1989, cheverud2001, siegmundyakir2007}. It is useful for clarity and full generality to separate the two distinct steps involved in this process, however.

This paper presents the details of the structural model outlined briefly in Section \ref{sec:intro}. By using mathematical notation for each abstraction, a framework with extraordinary explanatory power is devised. Section \ref{sec:theModel} provides an explanation of the model with all the necessary mathematical notation. The model is then used in a novel derivation of the Haldane \emph{map distance}, a common measure used to locate SNPs, in Section \ref{sec:derivingDists}. The utility of the model is further demonstrated in Section \ref{sec:correlation}, where it is used to derive the correlation between markers under classic genetic population settings. This results in an expression of the correlation between markers in any genetic study, which is simulated in Section \ref{sec:sim} under settings from the literature. Finally, Section \ref{sec:model2real} simulates the model and compares the results to theoretical expectations and experimental data in mice.

\section{A structural genetic model} \label{sec:theModel}

The structural model starts with
$$\m{G} = [\ve{g}_1| \ve{g}_2], \text{ } \ve{g}_1, \ve{g}_2 \in \mathcal{B}^{N_P}$$
where $\mathcal{B} = \{\text{adenine, guanine, cytosine, thymine}\}$ is the set of nucleotide bases and $N_P$ is the length of the genome. In humans $N_P \approx 3,234,830,000$. $\m{G}$ represents the whole genome of an individual, with all chromosomes placed sequentially in two adjacent columns corresponding to the maternally and paternally inherited variants. Though both of these variants are complete double-stranded sequences of DNA, nucleotides pair uniquely. Adenine binds exclusively with thymine and guanine binds exclusively with cytosine. Therefore $\ve{g}_1$ and $\ve{g}_2$ record the pattern only for one of the two DNA strands for each column, the complementary strand is implied by this sequence and the unique binding of nucleotides.

Rather than address the whole genome, GWAS typically deal with a selected subset of segments of interest. This is represented by
$$\m{S} = [\ve{s}_1 | \ve{s}_2], \text{ } \ve{s}_1, \text{ } \ve{s}_2 \in \mathcal{B}^K$$
with $K \ll N_P$. The mapping $\m{G} \rightarrow \m{S}$ chooses $K$ rows of $\m{G}$ to create $\m{S}$. This mapping is very seldom a random one. Previous work and databases of SNPs or other known markers motivate the choice of rows. Most commonly, then, the mapping $\m{G} \rightarrow \m{S}$ is a non-random selection of $M < K$ disjoint sequences from $\m{G}$.

In the case where $\m{S}$ contains only SNPs, the markers are most often \textit{biallelic}, i.e. the population is dominated by two different sequences or \textit{alleles} at the marker. These can be denoted using two different letters, such as $A$ and $B$, or analogously the uppercase and lowercase version of the same letter, such as $A$ and $a$. Converting the measured markers to letters is called annotation, a mapping $\m{S} \rightarrow \m{T}$ with
$$\m{T} = [\ve{t}_1 | \ve{t}_2], \text{ } \ve{t}_1, \text{ } \ve{t}_2 \in \{A,a\}^M.$$
Denoting the $i^{\text{th}}$ position of $\ve{t}_j$ as $t_{ij}$, $t_{lj} = A$ and $t_{mj} = A$ do not represent identical sequences at positions $l$ and $m$. Instead this indicates that the sequences annotated by the capital at each position are present at their respective positions.

These annotated variants in $\m{T}$ might next be converted to a numeric form. This is a mapping $\m{T} \rightarrow \m{X}$ such that
$$\m{X} := [\ve{x}_1 | \ve{x}_2], \text{ } \ve{x}_1, \text{ } \ve{x}_2 \in \Reals^M.$$
Commonly this is even more restricted with $\ve{x}_j \in \{0,1\}^M$ where
\begin{equation} \label{eq:indicator}
x_{ij} = \begin{cases}
  1, & \text{ if } t_{ij} = A \\
  0, & \text{ if } t_{ij} = a
\end{cases},
\end{equation}
is an indicator of the presence of the allele denoted with a capital.

Finally, $\m{X}$ may be converted into a vector
$$\ve{z} \in \Reals^M$$
summarizing the individual's inherited variants. There are many common mappings $\m{X} \rightarrow \ve{z}$. The \textit{dominance mapping} takes $z_i = \max\{x_{i1}, x_{i2}\}$, the \textit{homozygous mapping} uses $z_i = \ind{x_{i1}}{x_{i2}}$, and the \textit{additive map} is $\ve{z} = \ve{x}_1 + \ve{x}_2$, where $\ve{x}_1$ and $\ve{x}_2$ are given according to Equation \ref{eq:indicator} and $\ind{x}{y}$ is the indicator function
\begin{equation*}\ind{x}{y} = \begin{cases}
  1, & \text{ if } x = y \\
  0, & \text{ otherwise}
\end{cases}.\end{equation*} The additive map gives $\ve{z} \in \{0,1,2\}^M$ so $z_i$ is equal to the count of copies of $A$ at the $i^{\text{th}}$ marker across both of an individual's inherited variants.

Figure \ref{fig:modelDiagram} displays this model, with descriptive names added to each mapping. In the first step, $\m{G} \rightarrow \m{S}$, segments of the genome are \textit{selected} to obtain the marker sequences or interest. The next step, $\m{S} \rightarrow \m{T}$, \textit{annotates} the chosen markers by indicating which of the common alleles is present for that marker. These annotations are then converted to numeric values, or \textit{encoded}, in the step $\m{T} \rightarrow \m{X}$. Finally, the matrix $\m{X}$ is \textit{summarized} into a vector $\ve{z}$ with some row-wise operation.

\section{Deriving map distance} \label{sec:derivingDists}

The structural model presented in Section \ref{sec:theModel} is useful beyond being a clear guide to GWAS data. With only a few assumptions, it shows the Haldane map distance to be a corollary of the structure of DNA and mechanics of inheritance as understood today. This is in contrast to the typical derivation of map distance, which is agnostic to the structure of the genome and instead is based on Haldane's original differential equation from \cite{haldane1919}. Examples include \cite{kosambi1943estimation} and \cite{xu2013principles}. The derivation of the Haldane map using the model of Section \ref{sec:theModel} is outlined here, starting with a simple sketch of sexual reproduction.

\subsection{Sexual reproduction} \label{subsec:crossingover}

Sexual reproduction is the recombination of the genomes of two parents to create offspring genetically distinct from both. A distinction must be made between reproductive or \emph{sex} cells, e.g. sperm, and somatic cells. While somatic cells contain two variants of the genome, sex cells contain only one. When two sex cells combine, each provides its own variant to the offspring that results. Inheritance is mediated by the creation of sex cells, which itself involves the random selection of variants contained within somatic cells by meiosis.

To track the parental variants which may be inherited, introduce two matrices to represent the maternal and paternal genomes of which $\m{G}$ is the offspring:
$$\m{M} = [\ve{m}_1| \ve{m}_2] \text{ and } \m{F} = [\ve{f}_1| \ve{f}_2],$$
where $\ve{m}_1, \ve{m}_2, \ve{f}_1, \ve{f}_2 \in \mathcal{B}^{N_P}$. Crudely, sexual reproduction is the construction of $\m{G}$ from one random column of $\m{M}$ and one random column of $\m{F}$. So, $\m{G}$ could be $[\ve{m}_1 | \ve{f}_2]$, for example.

The real mechanism is much more complex. During meiosis, the columns of $\m{M}$ and $\m{F}$ are perturbed. Rather than being inherited by $\m{G}$ in the same form as in $\m{M}$ and $\m{F}$, regions in $\ve{f}_1$ may swap with regions in $\ve{f_2}$ and the same may occur with $\ve{m}_1$ and $\ve{m}_2$. This occurs either due to the \emph{independent assortment of chromosomes} or due to the \emph{crossing over} of variants.

Independent assortment is a direct consequence of the structure of the genome in somatic cells. Each chromosome is a separate molecule and so when sex cells are created, the variant of one chromosome inherited by offspring is independent of other chromosomes inherited from the same parent. This means that either of the paternal and maternal variants of a chromosome is equally likely to be passed on regardless of which variant is passed on for another chromosome.

Additionally, these variants may not be inherited identically as they appear in $\m{M}$ or $\m{F}$. There is a chance that the variants in a parent physically cross over each other while separating to form sex cells. Occasionally, this crossing results in a swap of the entire chromosome on either side of the cross, creating two completely new variants.

\subsection{Modelling cross overs} \label{subsec:modelcrossing}

Both crossing over and the independent assortment of chromosomes occur within each parental genome independently of the other parent, and so only one of the two needs to be considered in modeling cross overs. Suppose it is $\m{M}$.

Start with the assumption that genetic recombination is totally independent between chromosomes. Specifically, chromosomes not only assort independently but crossing over occurs independently on each chromosome and will affect only that chromosome's variants. This assumption can be thought of as a slightly stronger version of independent assortment. Therefore consider a vector
$$\ve{h} \in \{1, 2, \dots, C\}^{N_P}$$
for $C \in \Naturals$ which denotes the chromosomal membership of each row of $\m{M}$. For simplicity, set $h_i \leq h_j$ for all $i \leq j$. In other words, all base pairs of a chromosome appear in adjacent rows with some specified ordering of the chromosomes. Assuming cross overs occur independently for each chromosome, a cross over in chromosome $c$, say, will affect only those rows of $\m{M}$ where $\ve{h} = c$. Start with the simplest case, where $\ve{h}$ is a vector of ones. This is the case of a single chromosome, which can be extended to the entire genome by considering every other chromosome in the same way.

For this single chromosome, consider a cross over beginning at the $i^{\text{th}}$ base pair. This means the two variants of the chromosome physically cross at the $i^{\text{th}}$ base pair. Assuming the variants are always perfectly aligned so that the $i^{\text{th}}$ position on one variant will match with the $i^{\text{th}}$ on the other, each variant is consquently separated into two parts: the part up to, but not including, the $i^{\text{th}}$ base pair, and the part from the $i^{\text{th}}$ base pair until the end. These two parts are then swapped between the variants, so that the first part of one variant forms a new chromosome with the second part of the other. Whenever a cross over is said to ``begin at index $i$'', it will refer to this sort of crossing: a swap of the columns for the first $i-1$ rows of $\m{M}$ (or $\m{F}$). Introduce an indicator vector
$$\ve{V} = \tr{(V_1, \dots, V_{N_P})}$$
where
\begin{equation} \label{eq:crossindicator}
V_i = \begin{cases}
  1 & \text{ if a cross over at base pair } i \text{ occurs}, \\
  0 & \text{ otherwise},
\end{cases}
\end{equation}
and define $\sv{\pi} = (\pi_1, \pi_2, \dots, \pi_{N_P})$ so that $\pi_i = P(V_i = 1)$. This can be done without loss of generality, as the order of cross overs in time does not affect the final chromosome. Any chromosome, offspring, or sex cell for which any cross overs have occurred is called \textit{recombinant}.

As we rarely sequence the entire genome of an individual's somatic and sex cells, we will seldom see $\m{M}$ and its recombinant forms. Instead, just as $\m{S}$ is derived from $\m{G}$, $\m{M}_S$ and $\m{F}_S$ are derived from $\m{M}$ and $\m{F}$ respectively. Swaps of the markers of $\m{M}_S$ and $\m{F}_S$ as inferred from $\m{S}$ are then used to estimate the number of sex cells containing recombinant chromosomes. The proportion of sex cells produced with such a swap is called the \textit{recombination rate} for the pair of markers.

However, the recombination rate for a pair of markers tells us nothing of how many cross over events occurred between them. Any odd number of events leads to a swap, while any even number will be undetectable. With this restricted view, the true count of indices $i$ for which $V_i = 1$ cannot be known, and hence the $\pi_i$ cannot be estimated individually.

\subsection{Simplifying assumptions} \label{subsec:simplify}

Fortunately, if the recombination of two particular markers on the genome is all that is of interest, estimating individual $\pi_i$ values is unnecessary. Consider two such marker positions, $j$ and $k$ with $j < k$, and note that cross overs beginning at any of $j+1, j+2, \dots, k-1, k$ all result in these positions being split between variants. For identifiability assume that $\pi_{j+1} = \pi_{j+2} = \cdots = \pi_{k-1} = \pi_k = \pi_{j:k}$. Let $N_c$ be a random variable counting the number of cross overs beginning in the interval $\{j+1,j+2,\dots,k-1,k\}$. Then
$$P(N_c = n_c) = {k - j \choose n_c} \pi_{j:k}^{n_c} (1-\pi_{j:k})^{k - j - n_c}$$
if cross overs occur independently. For convenience, let $r = k - j$ and $\pi = \pi_{j:k}$, which gives
\begin{equation} \label{eq:binomialDist}
  P(N_c = n_c) = {r \choose n_c} \pi^{n_c} (1-\pi)^{r - n_c},
\end{equation}
where $r$ is a unitless count of base pairs between positions $j$ and $k$.

Recall that $N_P \approx 3,234,830,000$ in humans. This large number of base pairs spread over the 23 human chromosomes means that $j$ and $k$ will typically be separated by a great number of base pairs, and so $r$ will be very large. Indeed, examples in \cite{nyholt2004}, \cite{Salyakina2005}, and \cite{Galwey2009} have thousands or tens of thousands of base pairs between marker locations. Therefore, consider the limit of this expression as $r \rightarrow \infty$:
$$\lim_{r \rightarrow \infty} P(N_c = n_c) = \lim_{r \rightarrow \infty} {r \choose n_c} \pi^{n_c} (1-\pi)^{r - n_c}.$$
At this point, a substitution can be made:
$$\pi = \frac{\beta d(j,k)}{r} := \frac{\beta d}{r},$$
with $\beta, d(j,k) \in \Reals$. This substitution reparametrizes the probability $\pi$ with a rate parameter, $\beta$, a distance measure, $d(j,k)$, and the $r$ base pairs separating $j$ and $k$. As the units of $\beta$ and $d$ will always result in a unitless product, their selection is arbitrary. Any distance $d$ can be chosen and will invoke a corresponding $\beta$. If physical distance, for example in angstroms, were used, then $\beta$ would correspond to a rate of cross overs per unit length. One could alternatively use $d(j,k)=k-j$ and use a rate per base pair. This flexibility gives a great deal of freedom to choose a convenient set of units for measurement or understanding.

The substitution also leads to a substantial simplification, as
\begin{eqnarray} \label{eq:poissonlim}
  & & \lim_{r \rightarrow \infty} {r \choose n_c} \left ( \frac{\beta d}{r} \right )^{n_c} \left ( 1-\frac{\beta d}{r} \right )^{r - n_c} \nonumber\\
  & &  \nonumber \\
  & = & \lim_{r \rightarrow \infty} \frac{r^{n_c} + O(r^{n_c-1})}{n_c!} \left ( \frac{\beta d}{r} \right )^{n_c} \left ( 1-\frac{\beta d}{r} \right )^{r - n_c} \nonumber\\
  & & \nonumber \\
  & = & \frac{(\beta d)^{n_c}}{n_c!} \lim_{r \rightarrow \infty} \frac{r^{n_c} + O(r^{n_c-1})}{r^{n_c}} \left ( 1-\frac{\beta d}{r} \right )^{r - n_c} \nonumber\\
  & & \nonumber \\
  & = & \frac{(\beta d)^{n_c}}{n_c!} e^{-\beta d} = \lim_{r \rightarrow \infty} P(N_c = n_c),
\end{eqnarray}
\noindent the Poisson limit approximation for the binomial distribution.

Recall that if $N_c$ is odd, it will result in a swap of markers $j$ and $k$ between variants, while if $N_c$ is even, there will be no swap in the chromosome passed on. Define the recombination probability $p_r(d)$, which gives the probability of observing a swap for positions $j$ and $k$ with distance $d(j,k) := d$ between them. Then $p_r(d)$ is given by a sum of all odd terms from Equation \ref{eq:binomialDist}. Taking the simplification of Equation \ref{eq:poissonlim} gives
\begin{eqnarray} 
  & & \sum_{l = 0}^{\infty} \frac{(\beta d)^{2l + 1}}{(2l + 1)!} e^{-\beta d} \nonumber \\
  & & \nonumber \\
  & = & e^{-\beta d} \sum_{l = 0}^{\infty} \frac{(\beta d)^{2l + 1}}{(2l + 1)!}  \nonumber \\
  & & \nonumber \\
  & = & e^{-\beta d} \left ( \frac{e^{\beta d} - e^{- \beta d}}{2} \right ) \nonumber \\
  & & \nonumber \\
  & = & \frac{1}{2} \left ( 1 - e^{-2 \beta d} \right ) = p_r(d). \label{eq:haldanemap}
\end{eqnarray}
A final substitution converts Equation \ref{eq:haldanemap} to a form familiar to researchers in genomics. Setting $\beta = \frac{1}{100}$ so that each each unit increase in $d$ corresponds to a 0.01 increase in the expected number of cross overs gives Haldane's formula for the \textit{map distance} in \textit{centiMorgans} or cM.

By accounting for the structure of the genome and making a number of simplifying assumptions, the model from Section \ref{sec:theModel} gives this classic result of genetics without any reference to the population-level differential equation used in its original derivation. Indeed, it indicates this population level differential equation is a direct consequence of the structure of the genome. This powerful derivation can be taken a step further to compute a simple expression for genetic correlation.

\section{Genetic correlation} \label{sec:correlation}

\cite{cheverud2001, LiJi2005, Galwey2009} all present results based on the \emph{correlation between markers}. Recall $\ve{z}$ as depicted in Figure \ref{fig:modelDiagram} and described in the beginning of Section \ref{sec:intro}. For these papers, the correlation between markers refers to the observed correlation matrix of the vector $\ve{z}$ in a particular population. While the motivation of these authors is adjustment for multiple dependent testing, the importance of correlation in defining linkage disequilibrium makes the correlation structure of the genome a matter of general interest. This matrix can be determined analytically using the model of Section \ref{sec:theModel} and results of Section \ref{sec:derivingDists}.

For clarity, let $\ve{z}$ indicate an instance of the random vector $\ve{Z} = \tr{(Z_1, Z_2, \dots, Z_M)}$. Let the random vector $\ve{Z}$ follow the distribution of the summarized values $\ve{z}$ in a particular population. This population may be real, as is the case when this modelling is used in practice, or purely hypothetical, as will be the case in the following analysis.

Return to the annotated matrix $\m{T}$ and consider two markers at row indices $j$ and $k$. Introduce $\ve{c}$, which is defined similarly to $\ve{h}$ earlier, but now indicates chromosomal membership for the markers in $\m{T}$ rather than the base pairs in $\m{G}$. As individual markers are not split over chromosomes, $\ve{c}$ is always unambiguously defined.

There are two cases. Either $j$ and $k$ are on the same chromosome, that is $c_j = c_k$, or they are not, and so $c_j \neq c_k$. If these markers are not on the same chromosome, the assumptions of Section \ref{subsec:modelcrossing} dictate that there will be no correlation between $Z_j$ and $Z_k$, as these markers will assort independently alongside their respective chromosomes. If they are on the same chromosome, let $d(j,k) = d$ be the distance between them measured in cM as in Equation \ref{eq:haldanemap}. Denote the alleles of $j$ with $A$ and $a$ respectively and use $B$ and $b$ analogously for $k$. Assume that the pairwise association of these markers in the population is of interest, i.e. that all other markers on this chromosome can be ignored in analysis. This setting creates a radically simplified $\m{T}$, with 2 rows rather than $M$ and taking the form
$$\m{T} = \begin{bmatrix}
  A & a \\
  b & B \\
\end{bmatrix},$$
where the letters placed above are merely demonstrative. A simplified version of $\m{X}$ follows immediately from this $\m{T}$. Consider
$$\m{X} = \begin{bmatrix}
  x_{j1} & x_{j2} \\
  x_{k1} & x_{k2} \\
\end{bmatrix},$$
with all entries in $\{0,1\}$. As was the case for $\ve{z}$, these lowercase entries are realizations of random variables $X_{rs}$, $r \in \{j,k\}, s \in \{1,2\}$. Then $\m{X}$ implies a $\ve{Z}$ of
$$\ve{Z} = \begin{bmatrix} Z_j \\ Z_k \end{bmatrix} = \begin{bmatrix}
  X_{j1} + X_{j2} \\
  X_{k1} + X_{k2} \\
\end{bmatrix}$$
under the additive map. Consider $Corr(Z_j, Z_k)$ for a population resulting from the sexual reproduction of two known parents.
The mechanics of sexual reproduction outlined in Section \ref{subsec:crossingover} and the genotype of the parents reproducing to create $\m{X}$ determine the distribution of $Z_j$ and $Z_k$.

From the matrices $\m{M}$ and $\m{F}$ introduced alongside sexual reproduction, take simplified, annotated forms of these matrices to represent the paternal and maternal encodings at $j$ and $k$. Explicitly,
\begin{equation} \label{eq:parEncs}
  \m{F}_X = \begin{bmatrix}
  f_{j1} & f_{j2} \\
  f_{k1} & f_{k2} \\
\end{bmatrix} \text{ and }
\m{M}_X = \begin{bmatrix}
  m_{j1} & m_{j2} \\
  m_{k1} & m_{k2} \\
\end{bmatrix},
\end{equation}
where all entries are once again in $\{0,1\}$. Assume that $\m{F}_X$ and $\m{M}_X$ are known.\footnote{There are theoretical populations where this is true such as the $F_2$ intercross, where $f_{j1} = m_{j1} = f_{k1} = m_{k1} = 1$ and $f_{j2} = m_{j2} = f_{k2} = m_{k2} = 0$.} Additionally, introduce the difference matrix
\begin{equation} \label{eq:diffMatrix}
  \sm{\Delta} = \begin{bmatrix}
    f_{j1} - f_{j2} & m_{j1} - m_{j2} \\
    f_{k1} - f_{k2} & m_{k1} - m_{k2} \\
  \end{bmatrix} := \begin{bmatrix}
    \delta_{jF} & \delta_{jM} \\
    \delta_{kF} & \delta_{kM}
  \end{bmatrix}.
\end{equation}
This matrix will be useful in representing the correlation between $Z_j$ and $Z_k$.
Finally, assume that the variation in $\ve{Z}$ results purely from recombination by crossing over and independent assortment.

Begin with the expectation of $\ve{Z}$. Assuming no preferential inheritance of either variant, $X_{j1}$ is equally likely to be either $f_{j1}$ or $f_{j2}$ and so takes a uniform distribution over these two possibilities. A similar logic for all other entries in $\m{X}$ applies, and so
\begin{eqnarray}
    E[\ve{Z}] & = & {\begin{bmatrix}
        E[X_{j1}] + E[X_{j2}] \\
        E[X_{k1}] + E[X_{k2}] \\
      \end{bmatrix}} \nonumber\\
    & & \nonumber\\
    & = & {\frac{1}{2}\begin{bmatrix}
        f_{j1} + f_{j2} + m_{j1} + m_{j2} \\
        f_{k1} + f_{k2} + m_{k1} + m_{k2} \\
      \end{bmatrix}}, \nonumber
\end{eqnarray}
from which it follows
\begin{eqnarray}
    Var(Z_j) & = & E[(X_{j1} + X_{j2})^2] - E[Z_j]^2 \nonumber\\
    & & \nonumber\\ 
    & = & \frac{1}{4} \Big[ (f_{j1} + m_{j1})^2 + (f_{j2} + m_{j1})^2 + (f_{j1} + m_{j2})^2 + (f_{j2} + m_{j2})^2  - (f_{j1} + f_{j2} + m_{j1} + m_{j2})^2 \Big]. \nonumber
\end{eqnarray}
This can be simplified to give
\begin{eqnarray} 
  Var(Z_j) & = & \frac{1}{4} \left [ (f_{j1} - f_{j2})^2 + (m_{j1} - m_{j2})^2 \right ] \nonumber \\
  & & \nonumber \\
  & = & \frac{1}{4} \left [ \delta_{jF}^2 + \delta_{jM}^2 \right ]. \label{eq:z1var}
\end{eqnarray}
Analogously,
\begin{equation} \label{eq:z2var}
  Var(Z_k) = \frac{1}{4} \left [ \delta_{kF}^2 + \delta_{kM}^2 \right ].
\end{equation}
Considering the covariance:
\begin{eqnarray} 
    Cov(Z_j, Z_k) & = & Cov(X_{j1} + X_{j2}, X_{k1} + X_{k2}) \nonumber\\
    & & \nonumber \\
    & = & Cov(X_{j1}, X_{k1}) + Cov(X_{j1}, X_{k2}) + Cov(X_{j2}, X_{k1}) + Cov(X_{j2}, X_{k2}). \label{eq:covstep1}
\end{eqnarray}
So the covariance is re-expressed as a sum of four terms, each of which can be considered in turn.

$Cov(X_{j1}, X_{k2})$ and $Cov(X_{j2}, X_{k1})$ can be evaluated immediately. Both of these terms measure the covariance between values on the diagonals of $\m{X}$, that is the covariance between the maternally and paternally donated variants of the genome inherited from $\m{F}$ and $\m{M}$, respectively. These covariances therefore measure the amount of \emph{inbreeding} in a population, that is the degree to which parents tend to have the same genotype. \cite{crowkimura1970intro} quantify these covariances with a coefficient $r$ for general populations. With known parents, as in this case, these diagonal values are independent of each other and therefore uncorrelated. This can be confirmed by tedious algebra. Explicitly, $Cov(X_{j1}, X_{k2}) = Cov(X_{j2}, X_{k1}) = 0$. 

$Cov(X_{j1}, X_{k1})$ and $Cov(X_{j2}, X_{k2})$ measure the covariance of encodings on the same variant, and so cannot be so easily dismissed. Instead, consider $Cov(X_{j1}, X_{k1})$ and expand:
$$Cov(X_{j1}, X_{k1}) = E[X_{j1} X_{k1}] - E[X_{j1}]E[X_{k1}].$$
The equal probabiliy of inheritance of variants gives $E[X_{j1}] = \frac{1}{2}(f_{j1} + f_{j2})$ and $E[X_{k1}] = \frac{1}{2}(f_{k1} + f_{k2})$. Next consider $E[X_{j1} X_{k1}]$.

There are four possible values of $X_{j1} X_{k1}$, corresponding to inheritance of either of the two parental variants with or without recombination. If no recombination occurs, an event with probability $1 - p_r(d)$, either $f_{j1} f_{k1}$ or $f_{j2} f_{k2}$ is inherited with equal probability. If a cross over between $j$ and $k$ leads to recombination, then either $f_{j1} f_{k2}$ or $f_{j2} f_{k1}$ is passed on with equal probability. Accounting for these four possibilities gives
\begin{eqnarray}
  E[X_{j1} X_{k1}] & = & (1 - p_r(d)) \left ( \frac{1}{2} f_{j1} f_{k1} + \frac{1}{2} f_{j2} f_{k2} \right ) + p_r(d) \left ( \frac{1}{2} f_{j1} f_{k2} + \frac{1}{2} f_{j2} f_{k1} \right ). \nonumber
\end{eqnarray}
Combining this with the expectations of $X_{j1}$ and $X_{k1}$ gives
\begin{eqnarray}
    Cov(X_{j1}, X_{k1}) & = & E[X_{j1} X_{k1}] - E[X_{j1}]E[X_{k1}] \nonumber\\
    & & \nonumber\\
    & = & (1 - p_r(d)) \left ( \frac{1}{2} f_{j1} f_{k1} + \frac{1}{2} f_{j2} f_{k2} \right ) + p_r(d) \left ( \frac{1}{2} f_{j2} f_{k1} + \frac{1}{2} f_{j1} f_{k2} \right ) - \frac{1}{4} (f_{j1} + f_{j2})(f_{k1} + f_{k2}) \nonumber\\
    & & \nonumber\\
    & = & \frac{1}{4} \left ( 1 - 2 p_r(d) \right ) \Big ( f_{j1}f_{k1} + f_{j2}f_{k2} - f_{j2}f_{k1} - f_{j1}f_{k2} \Big ) \nonumber\\
    & & \nonumber\\
    & = & \frac{ 1 - 2 p_r(d)}{4}  \delta_{jF} \delta_{kF}. \label{eq:covstep2}
\end{eqnarray}
The same logic gives
\begin{equation} \label{eq:covstep3}
   Cov(X_{j2}, X_{k2}) = \frac{ 1 - 2 p_r(d)}{4}  \delta_{jM} \delta_{kM}.
\end{equation}
We obtain the covariance of $Z_j$ and $Z_k$ by adding the above and Equation \ref{eq:covstep1}. Substituting Equations \ref{eq:covstep2} and \ref{eq:covstep3} and $Cov(X_{j1}, X_{k2}) = Cov(X_{j2}, X_{k1}) = 0$ gives
\begin{equation} \label{eq:cov}
  Cov(Z_j, Z_k) = \frac{1 - 2 p_r(d)}{4} \Big [ \delta_{jF} \delta_{kF} + \delta_{jM} \delta_{kM} \Big ]. 
\end{equation}
Finally, Equations \ref{eq:z1var}, \ref{eq:z2var}, and \ref{eq:cov} can be combined to determine the correlation:
\begin{eqnarray} \label{eq:precorr}
    & & \frac{Cov(Z_j, Z_k)}{\sqrt{Var(Z_j) Var(Z_k)}} \nonumber\\
    & & \nonumber\\
    & = & \frac{ (1 - 2 p_r(d)) \Big [ \delta_{jF} \delta_{kF} + \delta_{jM} \delta_{kM} \Big ] }{  \sqrt{ \big ( \delta_{jF}^2 + \delta_{jM}^2 \big ) \big ( \delta_{kF}^2 + \delta_{kM}^2 \big )}} \nonumber\\
    & & \nonumber \\
    & := & (1 - 2 p_r(d)) \gamma = Corr(Z_j, Z_k),
\end{eqnarray}
where
\begin{equation} \label{eq:gammaDef}
\gamma = \frac{ \Big [ \delta_{jF} \delta_{kF} + \delta_{jM} \delta_{kM} \Big ] }{  \sqrt{ \big ( \delta_{jF}^2 + \delta_{jM}^2 \big ) \big ( \delta_{kF}^2 + \delta_{kM}^2 \big )}}.
\end{equation}
So, the correlation is a product of $(1-2 p_r(d))$, which depends on the markers in question, and a factor $\gamma$, which depends on the known parents. An even simpler expression is obtained by substituting the Haldane recombination probability from Equation \ref{eq:haldanemap} in place of $p_r(d)$:
\begin{eqnarray} \label{eq:corrdist}
    Corr(Z_j, Z_k) & = & (1 - 2 p_r(d)) \gamma \nonumber\\
    & & \nonumber\\
    & = & \left ( 1 - 2 \left [ \frac{1}{2} \left ( 1 - e^{-2 \beta d} \right ) \right ] \right ) \gamma \nonumber\\
    & & \nonumber\\
    & = & \gamma e^{-2 \beta d},
\end{eqnarray}
and so using the Haldane map distance the correlation between $Z_j$ and $Z_k$ decays exponentially in $d(j,k)$ with an intercept $\gamma$ determined by the parents' annotated matrices. As the entries in $\m{M}_X$ and $\m{F}_X$ are all 0 or 1, the differences in $\sm{\Delta}$ are all -1, 0, or 1. There are therefore $3^4 = 81$ potential $\gamma$ values, though most of these are not unique. 17 of these are undefined, corresponding to cases where $Var(Z_j) = 0$ or $Var(Z_k) = 0$. Table \ref{tab:gammaSum} summarizes the frequency of different $\gamma$ values for the remaining 64 combinations.
\begin{table}[!h]
  \begin{center}
  \begin{tabular}{rr} \hline
    $\gamma$ & Frequency \\ \hline
    $-1$ & 8 \\
    $-\frac{1}{\sqrt{2}}$ & 16 \\
    0 & 16 \\
    $\frac{1}{\sqrt{2}}$ & 16 \\
    1 & 8 \\ \hline
  \end{tabular}
  \caption{Frequency of $\gamma$ values across the 64 combinations for which correlation is defined} \label{tab:gammaSum}
  \end{center}
\end{table}
Only five symmetrically-distributed values are possible. A number of population settings for $\gamma$ are of particular interest due to their use in mouse breeding experiments \cite{green1966}. These will be outlined individually.

The first of these is the the \textit{$F_2$ intercross} design. \emph{Cross} here is short for sexual reproduction, rather than crossing over. This design considers the population resulting from the cross of $\m{M}_X$ and $\m{F}_X$ with
$$\m{F}_X = \m{M}_X = \begin{bmatrix}
  1 & 0 \\
  1 & 0 \\
\end{bmatrix}.$$
In this setting all the differences in $\gamma$ are 1 and so $\gamma_{\text{inter}} = 1$.

The next is the \textit{$N_2$ backcross}. Here the cross is between $\m{M}_X$ and $\m{F}_X$ defined as
$$\m{F}_X = \begin{bmatrix}
  f & f \\
  f & f \\
\end{bmatrix}, \text{ and }
\m{M}_X = \begin{bmatrix}
  1 & 0 \\
  1 & 0 \\
\end{bmatrix}.$$
where $f \in \{0,1\}$. In this setting, both differences defined on $\m{F}_X$ are 0 while both of those defined on $\m{M}_X$ are 1. This gives $\gamma_{\text{back}} = 1$, the same as that of the intercross population.

Other interesting cases without historical basis involve
$$\m{F}_X = \begin{bmatrix}
  0 & 1 \\
  1 & 0 \\
\end{bmatrix} \text{ or } \m{M}_X = \begin{bmatrix}
  0 & 1 \\
  1 & 0 \\
\end{bmatrix},$$
as these can result in $\gamma < 0$, and so a negative correlation. For example, if
$$\m{F}_X = \m{M}_X = \begin{bmatrix}
  0 & 1 \\
  1 & 0 \\
\end{bmatrix},$$
then $\gamma = -1$, while taking
$$\m{F}_X = \begin{bmatrix}
  0 & 1 \\
  1 & 0 \\
\end{bmatrix} \text{ and } \m{M}_X = \begin{bmatrix}
  0 & 1 \\
  0 & 1 \\
\end{bmatrix},$$
gives $\gamma = -\frac{1}{\sqrt{2}}$. Many other settings lead to no measured correlation. Take
$$\m{F}_X = \begin{bmatrix}
  0 & 1 \\
  1 & 0 \\
\end{bmatrix} \text{ and } \m{M}_X = \begin{bmatrix}
  1 & 1 \\
  0 & 0 \\
\end{bmatrix},$$
or
$$\m{F}_X = \begin{bmatrix}
  0 & 1 \\
  1 & 1 \\
\end{bmatrix} \text{ and } \m{M}_X = \begin{bmatrix}
  0 & 0 \\
  1 & 0 \\
\end{bmatrix},$$
for example. Note that these negative values are somewhat arbitrary. The encoding of 1 or 0 for particular alleles at a marker is not prescribed, but is rather an analytical choice. Therefore in any of these cases the encoding could be switched to give a positive $\gamma$ of the same magnitude. 

Finally, these results can be extended to the whole genome. Recalling that $j$ and $k$ were restricted to be markers on the same chromosome, this pairwise result can be generalized to the correlation matrix of $\ve{Z}$ for markers measured on different chromosomes. For markers on the same chromosome correlations will be proportional to $1 - 2p_r(d)$, where $p_r(d)$ is the probability of recombination as a function of the distance between markers. Based on the independent assortment of different chromosomes, the correlations will be zero for any pair $j$ and $k$ not on the same chromosome.

In other words, if $c_j = c_k$, Equation \ref{eq:precorr} dictates the correlation between $Z_j$ and $Z_k$. On the other hand, if $c_j \neq c_k$ the correlation between $Z_j$ and $Z_k$ will be zero. This implies a block diagonal structure corresponding to the chromosomes with correlations dictated by the probability of recombination within each chromosome. Most generally
\begin{equation} \label{eq:zcorr_gen}
  Corr(Z_j, Z_k) = \ind{c_k}{\{c_j\}} \gamma (1 - 2p_r(d)),
\end{equation}
and under the assumptions leading to the Haldane model Equation \ref{eq:corrdist} gives
\begin{equation} \label{eq:zcorr}
  Corr(Z_j, Z_k) = \ind{c_k}{\{c_j\}} \gamma e^{-2 \beta d(j,k)}.
\end{equation}

\section{Simulating the model} \label{sec:sim}

The results of Equation \ref{eq:zcorr} are simulated by combining the model in Section \ref{sec:theModel} with the map distance derivation of Equation \ref{eq:haldanemap}. A structure which mirrors $\m{T}$ is first created. It consists of two columns of annotated biallelic markers which may be on separate chromosomes. Intra-chromosome distances for those on the same chromosome are specified together with a function to generate recombination probabilities given these distances. By default, these distances are cMs and probabilities are given by Equation \ref{eq:haldanemap}.

A populaton can be generated from a pair of these matrices. For each individual offspring in the populuation, a few steps occur. First, cross overs are simulated using independently drawn Bernoulli random variables with probabilities given by the distances between markers. If a cross over occurs all intra-chromosomal rows of the parental matrix are swapped above the cross over index. Next the chromosomes of each offspring are selected from each parent independent of the selection of other chromosomes or the other parent. Simulating in this way creates dynamics consistent with the model in Section \ref{sec:theModel}. Each individual genome generated can then be encoded and summarized before the population-wide correlation matrix is computed.

Previous literature motivates the particular simulation settings used here. \cite{cheverud2001} investigates the correlation between markers by simulating a single chromosome with equidistant markers. All combinations of chromosome lengths of 50, 75, and 100 cM with markers equidistant at 50, 25, 12.5, and 6.25 cM were simulated for populations of 500 $F_2$ intercross offspring. \cite{LanderBotstein1989} instead simulates twelve chromosomes of length 100 cM with markers every 20 cM along each for a population of 250 $N_2$ backcross offspring.

Departing from a reference to distances in cM or base pairs, \cite{LiJi2005} set their simulation scenarios using the genetic $r^2$ measure as defined in \cite{hillrobertson1968}, which is exactly Pearson's product moment correlation for a two-by-two contingency table.
This difference is meaningful, as \cite{siegmundyakir2007} note that $r^2$ is not constant over generations. After $k$ generations it is given by
$$r_k^2 = \left [ 1 - p_r \right ]^{2k}  r^2_0$$
for two markers with $r^2 = r^2_0$ initially and a probability of recombination of $p_r$. Unlike cM or base pairs, which are constant over generations, $r^2$ eventually goes to zero.

Nonetheless, \cite{LiJi2005} simulate 10 independent regions within each of which 5 markers are placed such that adjacent markers have an $r^2$ of 0.8 between them. This design is analogous to that of \cite{LanderBotstein1989}, despite the difference in description.

\begin{figure}[h!]
  \begin{center}
    \begin{tabular}{cc}
      \includegraphics[scale=0.4]{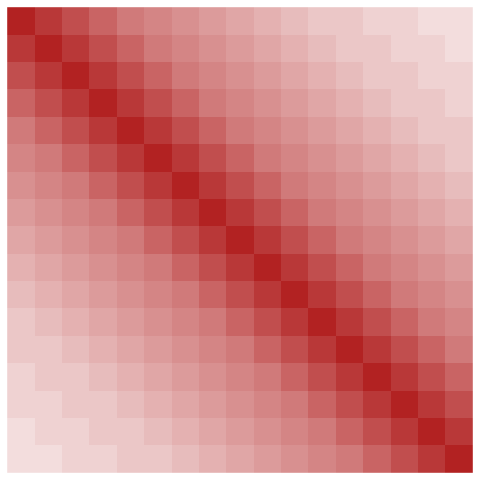} &
      \includegraphics[scale=0.4]{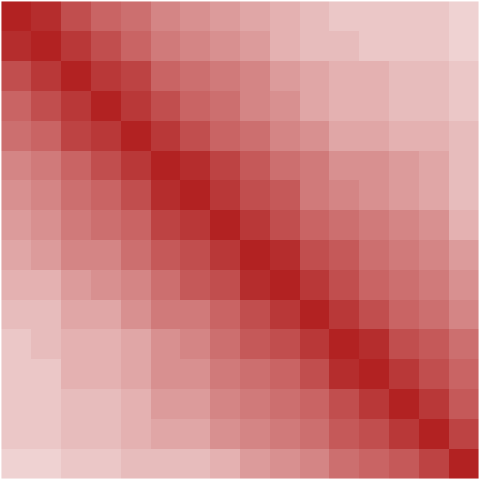} \\
      {\footnotesize (a) Theoretical} &
      {\footnotesize (b) Simulated} \\
    \end{tabular}
  \end{center}
  \caption{The correlation matrices of a population of 500 $F_2$ intercross offspring measured on a 100 cM chromosome with markers each 6.25 cM apart.}
  \label{fig:chevSims}
\end{figure}

\begin{figure}[h!]
  \begin{center}
    \begin{tabular}{cc}
      \includegraphics[scale=0.4]{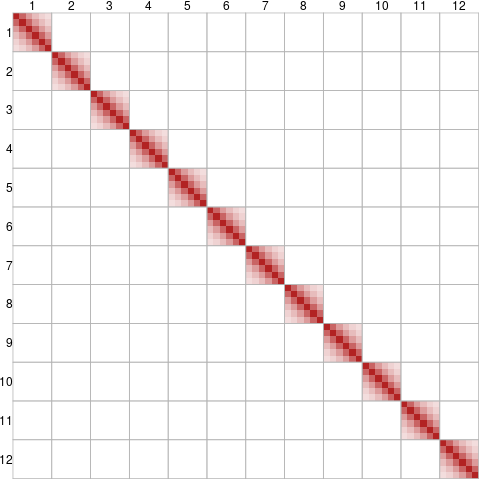} &
      \includegraphics[scale=0.4]{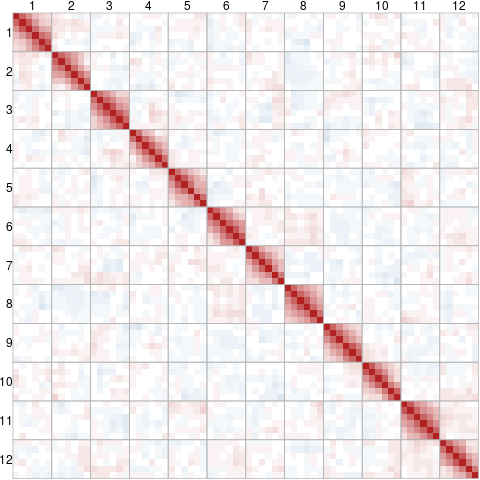} \\
      {\footnotesize (a) Theoretical} &
      {\footnotesize (b) Simulated} \\
    \end{tabular}
  \end{center}
  \caption{The correlation matrices of a population of 250 $N_2$ backcross offspring measured on twelve 100 cM chromosomes with markers 20 cM apart on each.}
  \label{fig:LBSims}
\end{figure}

The simulations of \cite{cheverud2001} and \cite{LanderBotstein1989} were recreated using the implementation detailed above. Specifically, these were the 100 cM chromosome with 6.25 cM separated markers of \cite{cheverud2001} and the twelve 100 cM chromosomes with 20 cM separated markers of \cite{LanderBotstein1989}. The resulting simulated correlation matrices and theoretical correlation matrices are visualized side by side using heatmaps in Figures \ref{fig:chevSims} and \ref{fig:LBSims}. In each heat map, the position of a square corresponds to the position of the corresponding correlation in the correlation matrix, and darker squares have a larger magnitude than lighter squares. Blue squares indicate negative correlation while red squares indicate a positive correlation.

Figure \ref{fig:chevSims}(a) displays a pattern of constant off-diagonal lines of decreasing value, as expected from Equation \ref{eq:haldanemap}. Roughly the same pattern is seen in Figure \ref{fig:chevSims}(b), though it is noisier. Rather than having clear constant lines along each off-diagonal, Figure \ref{fig:chevSims}(b) has regions of similar values which occur across several off-diagonal lines. This leads to the appearance of large squares of more strongly related values, a pattern absent from Figure \ref{fig:chevSims}(a).

Figure \ref{fig:LBSims} displays the setting of \cite{LanderBotstein1989} with the addition of guide lines to aid in reading the plot. These guide lines extend from labels on the top and left sides of the plot to indicate the chromosome of each marker. These lines create square borders along the diagonal which distinguish the intra-chromosome correlations within their borders from the inter-chromosome correlations outside of them. As suggested by Equation \ref{eq:haldanemap} and shown in Figure \ref{fig:LBSims}(a), Figure \ref{fig:LBSims}(b) has a stark block diagonal structure which agrees with these diagonal squares. The simulation therefore agrees very well with theory in this aspect. Within the chromosomes, there is also good agreement between Figure \ref{fig:LBSims}(a) and Figure \ref{fig:LBSims}(b). Both have decreasing correlations along the off-diagonal lines, with Figure \ref{fig:LBSims}(b) displaying similar departures from Figure \ref{fig:LBSims}(a) as Figure \ref{fig:chevSims}(b) does from Figure \ref{fig:chevSims}(a).

A more interesting noise pattern is seen between chromosomes outside the blocks in Figure \ref{fig:LBSims}(b). Unlike the strictly positive correlations seen in Figure \ref{fig:chevSims}, both negative and positive correlations are observed. Though many chromosomes show consistent patterns between their markers, with all correlations either positive or negative as between chromosomes 9 and 6 or 11 and 12, many have more complicated relationships. Between chromosomes 7 and 3, for example, both negative and positive correlations are observed between markers which are larger than the smallest intra-chromosomal correlations within 2.

\subsection{Implementation}

In order to make the above procedures repeatable, easily read, and flexible, the code for these simulations was implemented in \R using the S3 class system. The core construct was a class meant to reflect $\m{X}$.

Objects of the \code{genome} class are lists with two elements: \code{alleles} and \code{dists}. The \code{alleles} element is a list of two column matrices, where each matrix represents the value of $\m{X}$ for a particular chromosome. \code{dists} is a list of vectors of the same length as \code{alleles}. Each vector in \code{dists} gives the distances between the alleles in the corresponding element of \code{alleles}. A function called \code{abiogenesis} allows for the convenient creation of a genome through the specification of distances and allele values.

The function \code{sex} is then used to cross any pair of \code{genome} objects with the use of a \code{meiosis} helper function. \code{sex} accepts an arbitrary distance function which is passed to \code{meiosis} to convert the \code{dists} elements of the \code{genome}s into probabilities of crossing over. By default, the Haldane map distance conversion of Equation \ref{eq:haldanemap} is used. Random Bernoulli trials for each of these probabilities then determines the locations of cross over events, and sections of the columns of \code{alleles} are swapped accordingly.

The correlation of repeated swaps is then computed via \code{popCorrelation}. As the choice of scoring is at the discretion of the analyst, it accepts not only a list of \code{genome} objects representing the result of repeated crosses, but also a scoring function which accepts a single \code{genome} and returns a $\ve{z}$ value. By default, the additive scoring of $\ve{z} = \ve{x}_1 + \ve{x}_2$ is used.

The score function generates the $\ve{z}$ values for each \code{genome} provided to \code{popCorrelation}, and correlations between the elements of these $\ve{z}$ are computed. Finally, the \code{image} wrapper \code{corrImg} displays a correlation matrix rearranged so that the main diagonal is consistent with the typical arrangement of correlation matrices. 

\section{Comparing the model to reality} \label{sec:model2real}

Though simulation confirms that population correlations generated under the model of Section \ref{sec:theModel} match the predictions of Equation \ref{eq:zcorr}, the true test of any model must involve empirical measurements. This requires data other than in \cite{LanderBotstein1989} and \cite{cheverud2001}, who only perform simulations.

The Mouse Genome Database (MGD) \cite{bultetal2019mouse} provides annotation data for more than a dozen mouse populations resulting from crosses of known breeds or \emph{strains}. The database also provides the references needed to determine cM distances between markers measured in these experiments. All of these resources are publicly provided at the Mouse Genome Informatics website: \href{www.informatics.jax.org}{www.informatics.jax.org}.

Considering only those experiments with complete observations leaves several data sets. Two of these investigate an identical population setting: the \emph{BSB mouse cross} of \cite{fisleretal1993bsb}. BSB mice are those resulting from the $N_2$ backcross of the C57BL/6J and {\it Mus Spretus} mouse strains, detailed respectively in \cite{C57BL6J} and \cite{dejageretal2009mspretus}. The first of these crosses is the \emph{JAX BSB} cross of \cite{roweetal1994jaxbsb} and the second is the \emph{UCLA BSB} cross of \cite{welchetal1996uclabsb}. Both the JAX and UCLA BSB cross data were downloaded from the \href{www.informatics.jax.org/downloads/reports/index.html}{the Mouse Genome Database}.

\begin{figure}[h!]
  \begin{center}
    \begin{tabular}{ccc}
      \includegraphics[scale = 0.26]{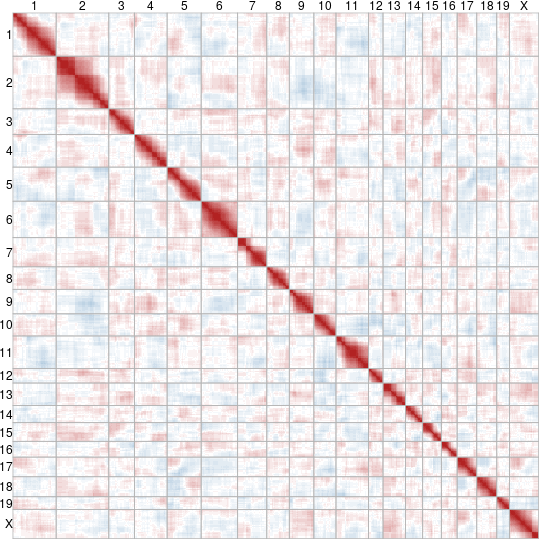} & \includegraphics[scale = 0.26]{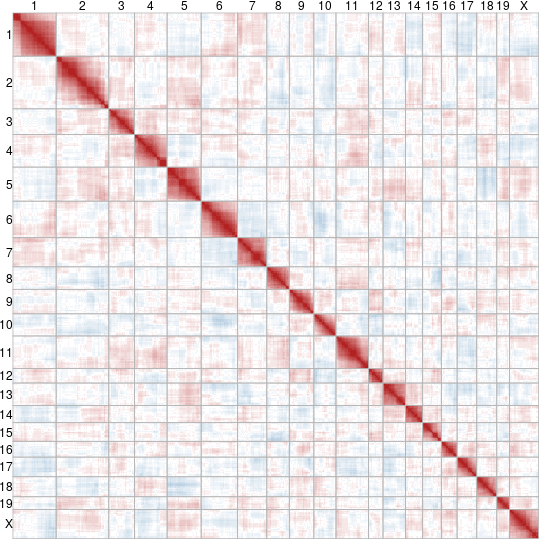} & \includegraphics[scale = 0.26]{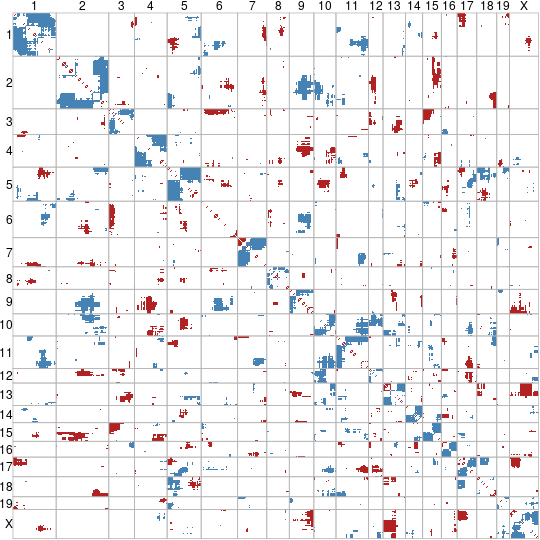} \\
      {\footnotesize (a) Experiment} & {\footnotesize (b) Simulated example} & {\footnotesize (c) Observed quantiles}
    \end{tabular}
  \end{center}
  \caption{Observed and simulated correlations for the \href{http://www.informatics.jax.org/downloads/reports/MGI_JAX_BSB_Panel.rpt}{MGD JAX BSB cross} from \cite{roweetal1994jaxbsb}. (c) displays quantiles determined from 10,000 simulated crosses, those quantiles less than 250 are shaded blue and those greater than 9,750 are shaded red.}
  \label{fig:jaxbsb}
\end{figure}

\begin{figure}[h!]
  \begin{center}
    \begin{tabular}{ccc}
      \includegraphics[scale = 0.26]{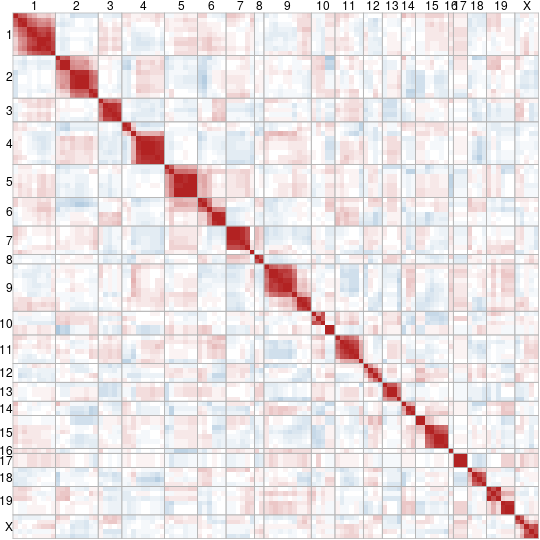} & \includegraphics[scale = 0.26]{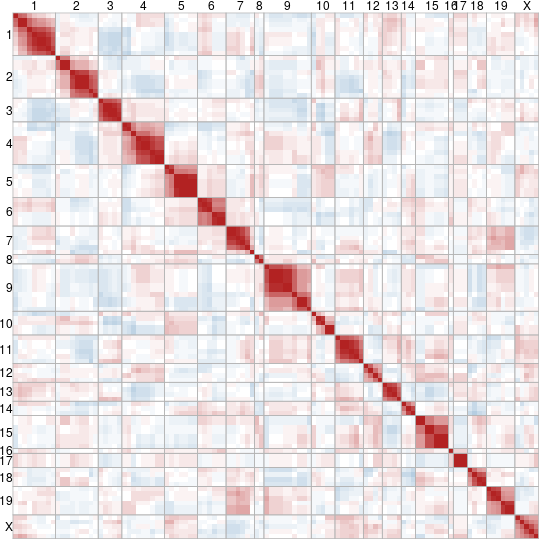} & \includegraphics[scale = 0.26]{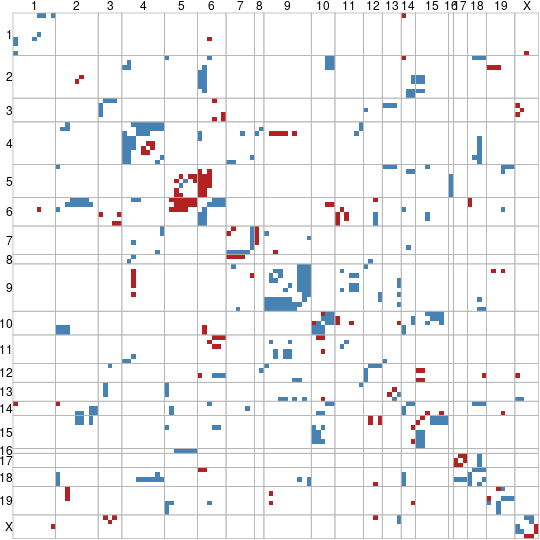} \\
      {\footnotesize (a) Experiment} & {\footnotesize (b) Simulated example} & {\footnotesize (c) Observed quantiles}
    \end{tabular}
  \end{center}
  \caption{Observed and simulated correlations for the \href{http://www.informatics.jax.org/downloads/reports/MGI_UCLA_BSB_Panel.rpt}{MGD UCLA BSB cross} from \cite{welchetal1996uclabsb}. (c) displays quantiles determined from 10,000 simulated crosses, those quantiles less than 250 are shaded blue and those greater than 9,750 are shaded red.}
  \label{fig:uclabsb}
\end{figure}

The data sets require further cleaning before being used, however. First, any markers which do not have a known position along a chromosome in cMs are removed from both data sets. Following that, any individual mice with incomplete data are excluded. For the JAX BSB data this leaves 94 mice annotated at 1496 markers while the UCLA BSB data has 66 mice annotated at 111 markers. The correlation matrices for these data sets are displayed in Figures \ref{fig:jaxbsb}(a) and \ref{fig:uclabsb}(a) respectively.

To determine the expected distribution of these correlations, the cM positions of measured markers were used to simulate 10,000 crosses under each of the JAX BSB and UCLA BSB settings using the methods of Section \ref{sec:sim}. Figures \ref{fig:jaxbsb}(b) and \ref{fig:uclabsb}(b) display example correlation matrices from one such simulated population. For each setting, the quantile of the each experimental pairwise correlation was then computed using the 10,000 simulated crosses. Figures \ref{fig:jaxbsb}(c) and \ref{fig:uclabsb}(c) display those quantiles which are less than 250 in blue and those which are greater than 9,750 in red for their respective settings. These correspond to unadjusted two-sided 95\% confidence rejection regions for each correlation.

Qualitatively, the simulated examples show good agreement to experimental results. In both Figures \ref{fig:jaxbsb} and \ref{fig:uclabsb} the patterns of correlation between chromosomes are similar between experiment and simulation. Figures \ref{fig:jaxbsb}(c) and \ref{fig:uclabsb}(c) additionally suggest that these patterns are little more than noise. The regions of unusually strong correlations shaded in red do not appear to follow any clear pattern, nor do the patterns of unusually weak correlations shaded in blue.

The similarity continues within chromosomes. Figures \ref{fig:jaxbsb}(c) and \ref{fig:uclabsb}(c) are generally not shaded within chromosomes. In particular, very little of the region close to the diagonal is shaded. The most noteworthy pattern in either sub-plot occurs in the corners of the diagonal squares indicating chromosomes in Figure \ref{fig:jaxbsb}(c). Many of these corners are shaded blue, suggesting these distant intra-chromosome correlations are less than might be expected. The pattern of shading is suggestive of block structures within chromosomes where contiguous sections are fit well by the model but may have more complex dynamics between them. 

A likely explanation is the non-independence, or \emph{interference}, of cross overs. \cite{bromanetal2002crossover} evaluated the pattern of cross overs in the cross of \cite{roweetal1994jaxbsb}, the basis of the JAX BSB data. Their results suggest that cross overs are not fully independent. Most mouse chromosomes are much less than 100 cM in length, yet cross overs rarely occur within 20 cM of each other and fewer cross overs than expected tend to occur on the same chromosome. This interference will have little impact on the correlation between markers with short distances between them, as more than one cross over event is unlikely to occur in a short interval. Markers separated by longer distances are impacted by this observed interference to a greater extent, as the observed number of double cross overs will be less than expected. This increases the chance that distant markers will be separated in meiosis by a single crossover, leading to a weaker correlation than predicted.

\begin{figure}[h!]
  \begin{center}
    \begin{tabular}{cc}
      \includegraphics[scale = 0.35]{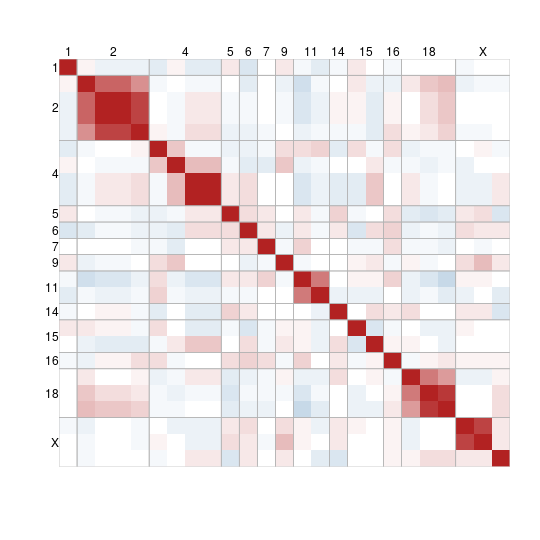} &
      \includegraphics[scale = 0.35]{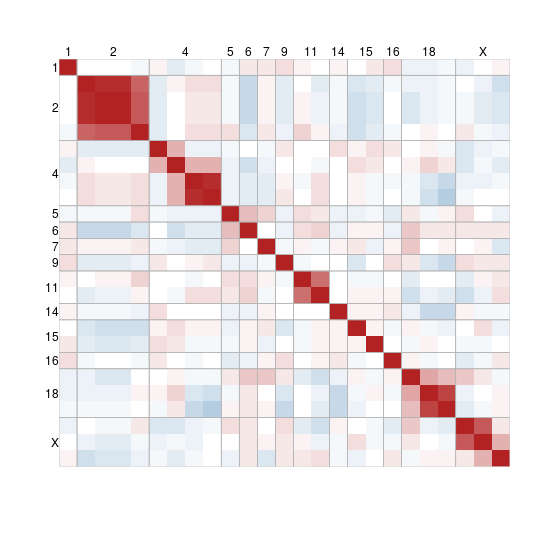} \\
      {\footnotesize (a) JAX BSB data} &
      {\footnotesize (b) UCLA BSB data}
    \end{tabular}
  \end{center}
  \caption{Pairwise correlations for the common marker positions of the JAX and UCLA BSB data.}
  \label{fig:bsbcommon}
\end{figure}

That said, it is important not to over-interpret this pattern. It is not repeated in Figure \ref{fig:uclabsb} and the shading of quantiles has not been adjusted to account for the many multiple tests performed in each plot. In order to get a greater sense of this experimental departure from theory, the common markers measured between the UCLA BSB and JAX BSB data were identified and the correlation matrices computed for these common locations. Having two experimental replicates should reduce the impact of one unusual measurement. These correlations are displayed in Figure \ref{fig:bsbcommon}. Most chromosomes have only one marker measured in common between these experiments, but chromosomes 2, 4, and 18 have several.

These common markers motivated further simulation. The common markers on chromosomes 2, 4, and 18 were used to generate 10,000 simulated crosses of 80 mice, the average of the JAX and UCLA BSB cross populations. Correlation matrices were computed for each population. Additionally, 10,000 paired populations of each of the JAX and UCLA crosses were simulated independently, and the average of these correlation matrices computed for each pair. Independence was assumed because the experiments of \cite{roweetal1994jaxbsb} and \cite{welchetal1996uclabsb} were carried out years apart in different labs.

\begin{figure}[h]
  \begin{center}
    \begin{tabular}{ccc}
      \includegraphics[scale = 0.5]{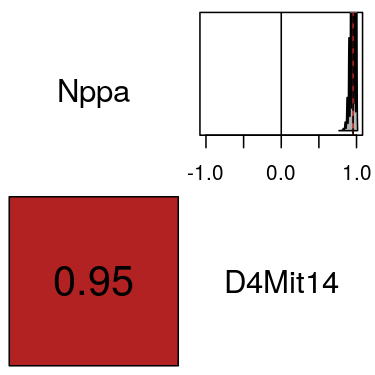} & \includegraphics[scale = 0.5]{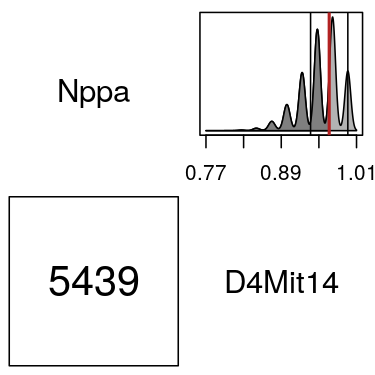} & \includegraphics[scale = 0.5]{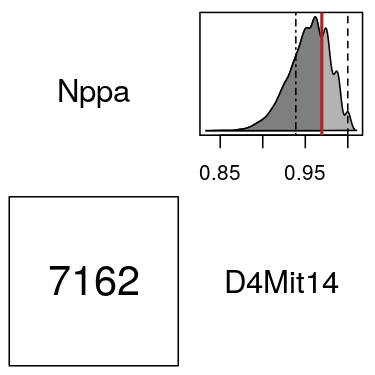} \\
      {\footnotesize (a) Correlation and distribution} & {\footnotesize (b) Mean quantile and distribution} & {\footnotesize (c) Mean quantile and mean distribution}
    \end{tabular}
  \end{center}
  \caption{The pairwise units of the correlation distribution and correlation test plots. (a) and (b) are based on 10,000 simulated populations of 80 individuals while (c) uses the means of 10,000 pairs of populations of 66 and 94.}
  \label{fig:2by2}
\end{figure}

For any pair of markers this gives two resulting distributions. One of the correlation over repeated crosses of an averaged population size and one showing the distribution of the correlations averaged across repeated crosses in two paired populations. Both of these are compared to the JAX and UCLA data using the two-by-two layouts of Figure \ref{fig:2by2}.
Each subplot of Figure \ref{fig:2by2} consists of four cells. Along the diagonal, each cell provides one MGD marker symbol to display the pair of markers being compared, in this case Nppa and D4Mit14. Above the diagonal a distribution is displayed, in all cases the kernel density estimate (KDE) of the distribution of interest. Below the diagonal a numeric summary with some informative shading is displayed.

The upper cell of Figure \ref{fig:2by2}(a) displays the KDE of the averaged population size data on a plot reflecting the range of possible correlations. A solid black line in this cell is drawn at the theoretical value predicted by Equation \ref{eq:haldanemap} and a dashed red line at the mean correlation across all simulations. The lower cell reports this mean shaded with the divergent palette of Figures \ref{fig:chevSims} to \ref{fig:bsbcommon}.

Figure \ref{fig:2by2}(b) displays this KDE in more detail by changing the bounds of the upper cell. Two solid lines are added at the correlations computed from the JAX and UCLA BSB data and a thicker red line is added at their mean, below which the density is shaded. The lower cell of Figure \ref{fig:2by2}(b) communicates the quantile of this red line, that is how many simulated correlations fall in the shaded region. As in Figures \ref{fig:uclabsb} and \ref{fig:jaxbsb}, this cell is shaded blue if the quantile is less than 250 and red if it is greater than 9,750.

Figure \ref{fig:2by2}(c) is identical to \ref{fig:2by2}(b), but with a more appropriate KDE. To model the distribution of the mean of the JAX and UCLA correlations for a pair of markers, 10,000 independent BSB crosses of both 66 and 94 mice were simulated. For each independent pair, the correlation is computed on both populations and averaged. This gives 10,000 realizations of the average correlation of the JAX and UCLA data under the model. Additional information is encoded by line type. The JAX correlation is marked with a dashed line while the UCLA data is marked with a dot-dashed line.

Figure \ref{fig:2by2} suggests the model fits well for Nppa and D4Mit4. Figure \ref{fig:2by2}(a) reiterates the close agreement of simulation and theory: the simulated mean and the theoretical value are both at 0.95. Across all simulations, the KDE shows universally strong, positive correlations above 0.5. The experimental values do not seem out of the ordinary range of these simulated correlations in Figure \ref{fig:2by2}(b), nor does their mean. This suggests the model works well for this pair. Figure \ref{fig:2by2}(c) reinforces this conclusion. The quantile of the mean is near the centre to the distribution of analogous means of paired, independent populations.

\begin{figure}[h]
  \begin{center}
      \includegraphics[scale = 0.5]{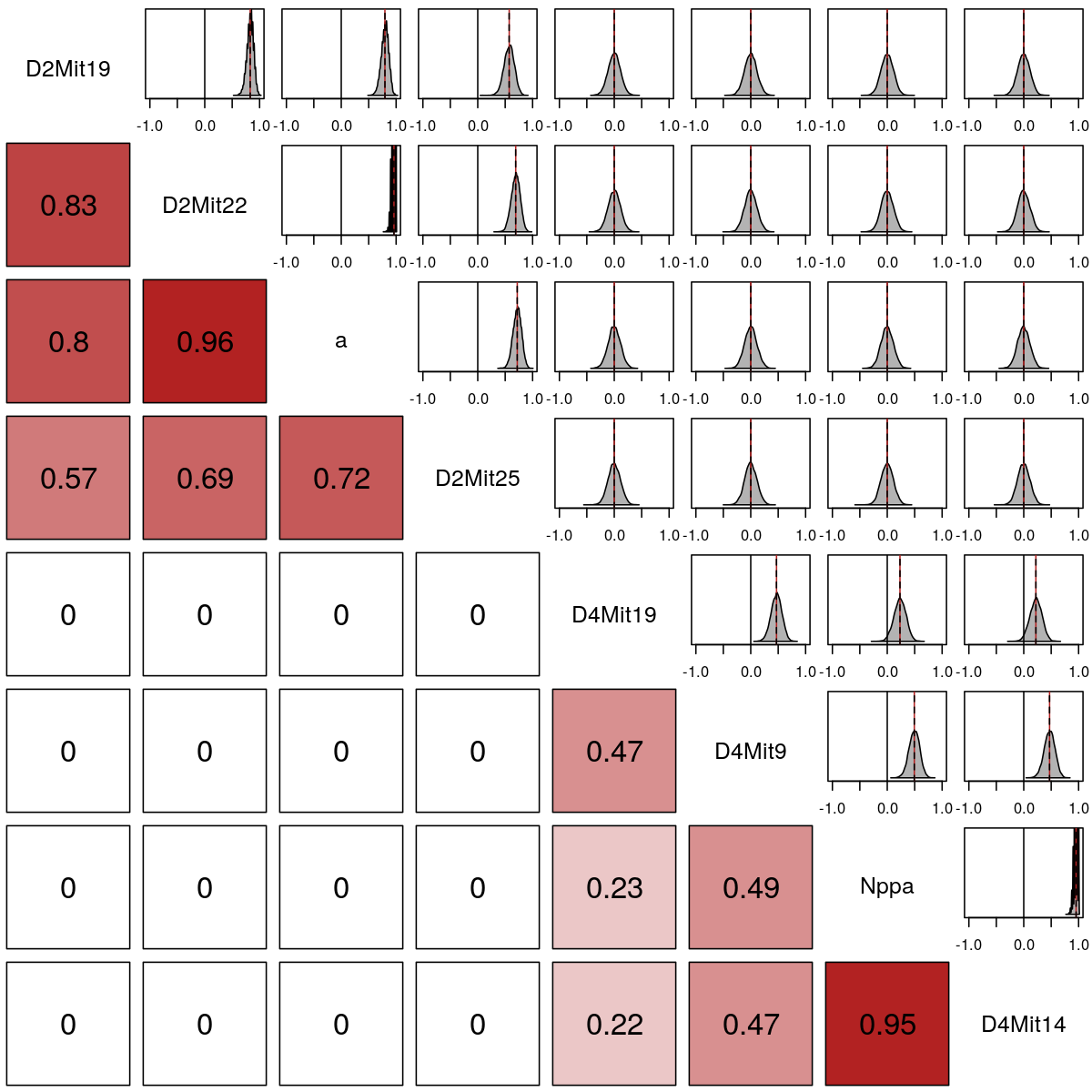}
  \end{center}
  \caption{The \emph{correlation distribution plot} for 10,000 simulated BSB crosses of the common markers on chromosomes 2 and 4 in the JAX and UCLA data.}
  \label{fig:bsbcorrDist}
\end{figure}

\begin{figure}[h]
  \begin{center}
      \includegraphics[scale = 0.5]{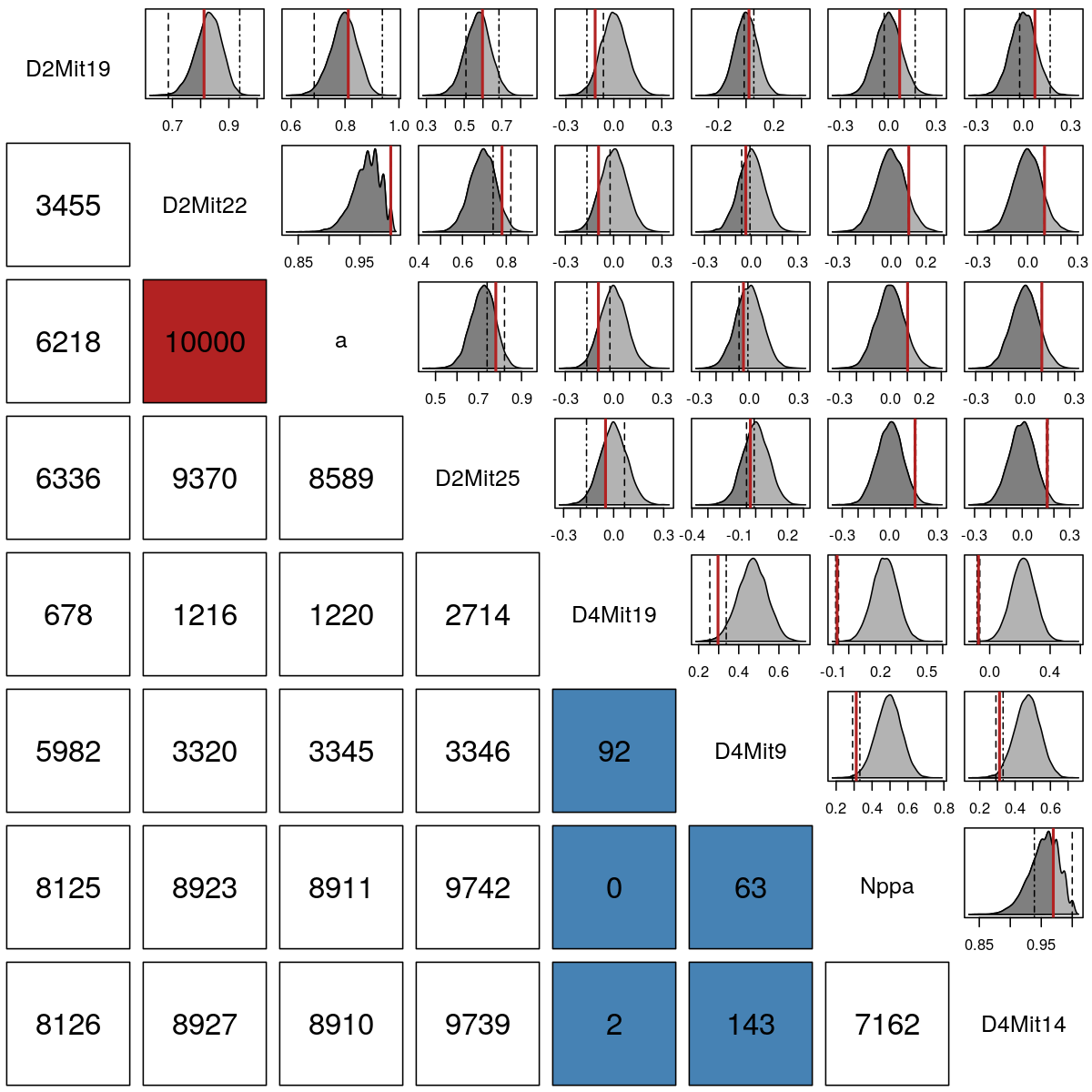}
  \end{center}
  \caption{The \emph{correlation test plot} for the JAX and UCLA BSB crosses. The upper cells show the distribution of 10,000 simulated averaged correlations between the JAX and UCLA BSB crosses. The experimental results are marked by black lines and their mean marked by a red line. The bottom cells give the quantile of the corresponding mean over the 10,000 simulated crosses.}
  \label{fig:bsbcorrTest}
\end{figure}

The subplots of Figure \ref{fig:2by2} are expanded into larger displays by adding more markers along the diagonal so the original subplots become pairwise units repeated over a larger array. Expanding to include all common markers on chromosomes 2 and 4 between the JAX and UCLA BSB data gives an eight by eight matrix. Along the diagonal, all eight marker symbols are displayed. For any cell, the corresponding pair of markers is found by tracing along its row and column until the diagonal is reached. Interpretation then follows as in the two-by-two case. In an array, the pattern across all markers can be seen at a glance, especially in the lower cells, which can then guide the inspection of specific pairs in more detail. Such an expansion of Figure \ref{fig:2by2}(a) gives the \emph{correlation distribution plot} of Figure \ref{fig:bsbcorrDist}, while for Figure \ref{fig:2by2}(c) the \emph{correlation test plot} of Figure \ref{fig:bsbcorrTest} results.

In Figure \ref{fig:bsbcorrDist}, the distributions of simulated correlations are generally symmetric and unimodal with centres at the theoretical correlation. Indeed, the means of simulated correlations agree with theory to two decimals for all pairwise correlations. The shape and spread of the distribution of correlations seems highly dependent on the proximity of a pair of markers. Markers which are close together in cM and have a high correlation display very little variation across the simulations relative to markers which are further apart on the same chromosome or are on different chromosomes. Additionally, the KDEs of the close markers have separate peaks and roughness indicative of clusters of correlation values. Figure \ref{fig:bsbcorrTest} shows this as well, but it has been smoothed somewhat by averaging.

Of the twenty eight lower cells of Figure \ref{fig:bsbcorrTest}, six are shaded. The first of these, between markers D2Mit22 and a, is perhaps misleading. The observed quantiles are computed by counting the values less than or equal to that observed, but this pair has an observed mean correlation of 1. It is therefore necessarily larger than or equal to all other mean correlations, despite having an identical value as 291 simulated means. This shading should therefore be ignored, as the value is not so unusual.

\begin{figure*}[htp]
  \begin{center}
      \includegraphics[scale = 0.5]{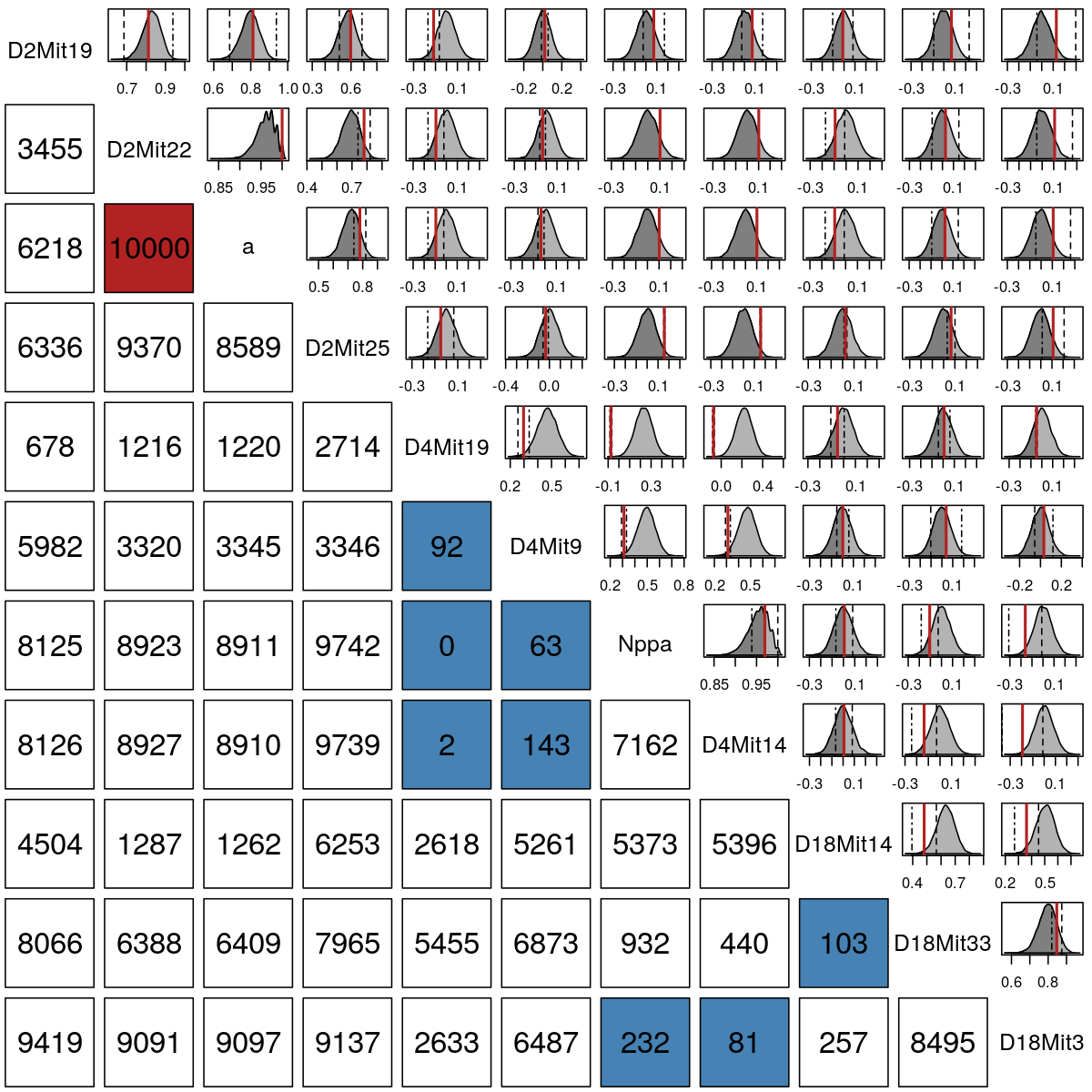}
  \end{center}
  \caption{Adding chromosome 18's three common markers to Figure \ref{fig:bsbcorrTest}.}
  \label{fig:bsbcorrTestBig}
\end{figure*}

All of the remaining shaded cells, with lower correlations than expected, involve chromosome 4 alone. Moreoever, the lines showing the independent experimental observations for these pairwise correlations in tbe JAX and UCLA data are much less variable than other cells. They are almost identical between experiments. This consistent departure of markers on chromosome 4 from the expectations of the model therefore suggests that chromosome 4 may experience stronger cross over intereference than chromosome one. This hypothesis is bolstered when chromosome 18 is added in Figure \ref{fig:bsbcorrTestBig}.

While it is harder to read the cells with chromosome 18 added, only one cell within chromosome 18 is shaded. The shaded pair, D18Mit14 and D18Mit33, does not show the close agreement over experiments as does chromosome 4. Chromosome 4 is therefore noteworthy for the poorer fit of the model to its correlations and the consistency of these correlations over independent experiments.

This demonstrates the value of the correlation test plot in Figure \ref{fig:bsbcorrTest} as a diagnostic. Repeated simulation under a model with no interference is used to create the distributions seen in the upper cells. Any departure from these distributions, especially a consistent departure over multiple experiments, shows where this model is inadequate. This suggests that these regions may have more cross over interference or generally more complex dynamics than others where the model fits well, and may be a priority for future measurement or research into more complicated inheritance patterns.

\subsection{The impact of cross overs on correlation} \label{subsec:corrCountCO}

\begin{figure}[htp]
  \begin{center}
    \begin{tabular}{cc}
      \includegraphics[scale = 0.5]{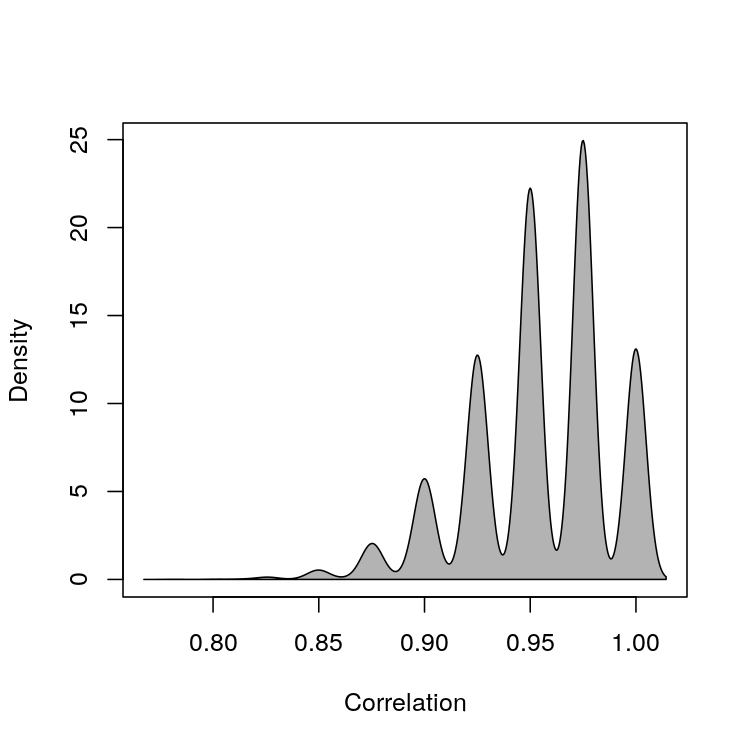} & \includegraphics[scale = 0.5]{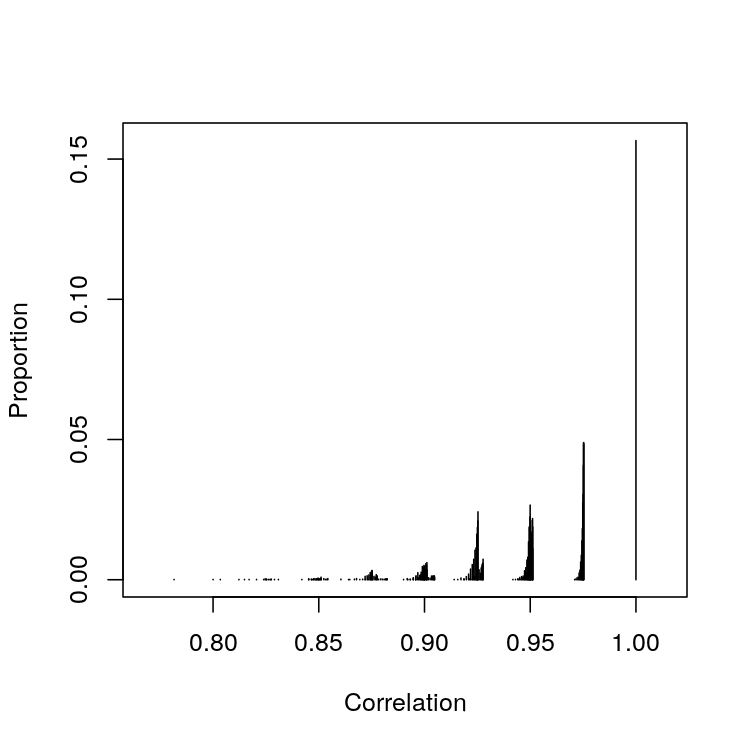} \\
      {\footnotesize (a) KDE} & {\footnotesize (b) Bar plot}
    \end{tabular}
  \end{center}
  \caption{The distribution of the correlation between markers Nppa and D4Mit14 over 10,000 simulated BSB crosses.}
  \label{fig:discrete}
\end{figure}

Another noteworthy feature is the roughness seen in Figure \ref{fig:2by2}(b) and repeated to a lesser extent in Figure \ref{fig:2by2}(c). A close inspection of the pairs D2Mit22/a and Nppa/D4Mit14 reveals that both seem to cluster around the same seven values across repeated simulation. This suggests the correlations for highly associated markers take several discrete values in a population. Figure \ref{fig:discrete} shows that the correlation distribution is much more complicated with a KDE and bar plot of correlations between Nppa and D4Mit14 across the 10,000 simulated populations of average size.

The bar plot in Figure \ref{fig:discrete}(b) places lines with lengths proportional to the number of simulated populations with a given correlation at the corresponding point on the horizontal axis. Seven clear clusters are visible, and a handful of lines suggest a potential eighth. The first of these from the right is at 1, and is a single line rather than a collection of lines. The peaks of each of the following groups of lines correspond to the peaks in Figure \ref{fig:discrete}(a). This intriguing pattern can be explained by the features of the backcross being simulated and the equation for sample correlation.

Recall that these correlations result from the simulation of the BSB cross, a specific backcross. In general, the backcross of the marker summaries $Z_j$ and $Z_k$ under the additive map can be modelled by the sexual reproduction of the annotated matrices
$$\m{F}_X = \begin{bmatrix}
  f & f \\
  f & f \\
\end{bmatrix} \text{ and }
\m{M}_X = \begin{bmatrix}
  1 & 0 \\
  1 & 0 \\
\end{bmatrix}.$$
Without loss of generality, take $f = 1$. In the absence of cross overs there are only two possible values for $\ve{z}$: $\tr{(1,1)}$ and $\tr{(2,2)}$. Both of these values of $\ve{z}$ are equally likely, as the inherited variants are independently donated by each parent. Regardless of which is inherited, $Z_{ij} = Z_{ik}$ for all of the offspring indexed by $i$ in a population of $n$ without cross overs. The observed sample correlation, given by
\begin{equation} \label{eq:sampleCorr}
\frac{\sum_{i = 1}^n z_{ij} z_{ik} - n \bar{z}_{j} \bar{z}_{k}}{\sqrt{ \left ( \sum_{i = 1}^n z_{ij}^2 - n \bar{z}_{j}^2 \right ) \left ( \sum_{i = 1}^n z_{ik}^2 - n \bar{z}_{k}^2 \right )}}
\end{equation}
where $\bar{z}_l = \frac{1}{n} \sum_{i = 1}^n z_{il}$, is therefore always 1.

Suppose $m$ cross overs occur and are inherited by the last $m$ offspring. Then there are $m$ recombinant individuals in the population with $\ve{z} = \tr{(1,2)}$ or $\ve{z} = \tr{(2,1)}$. Let $m_{12}$ be the count of $\tr{(1,2)}$ and $m_{21}$ be the count of $\tr{(2,1)}$. Take $Z_k$ as the \emph{reference marker} and assume the recombinant individuals would have taken that value at both markers without a cross over. That is, assume $\tr{(2,1)}$ individuals would have been $\tr{(1,1)}$ and $\tr{(1,2)}$ would have been $\tr{(2,2)}$ without recombination. Let
\begin{eqnarray} 
  r^2_{\text{ref}} & := & \frac{ s_{jk}^* - n \bar{z}_{j}^* \bar{z}_{k}^*}{\sqrt{ \big ( s_{jj}^* - n (\bar{z}_{j}^*)^2 \big ) \big ( s_{kk}^* - n (\bar{z}_{k}^*)^2 \big )}} \label{eq:refCorr} \\
  & = & \frac{\widehat{\sigma}^2_{jk}}{\sqrt{ \widehat{\sigma}^2_{jj} \widehat{\sigma}^2_{kk} }} = 1\nonumber
\end{eqnarray}
represent all quantities without recombination.
As there cannot be more cross overs than the corresponding number present in the reference marker
$$m_{12} \leq n(\bar{z}_k^* - 1) \text{ and}$$
$$m_{21} \leq n(2 - \bar{z}_k^*).$$
With $m = m_{12} + m_{21}$ recombinant individuals, the sums over $z_{ij}$ in Equation \ref{eq:refCorr} will change. Sums over $z_{ik}$ will not change, as these are the reference values.

The impact can be considered on a term-wise basis. $z_{ij}$ increases by 1 if recombination changes marker $j$ in individual $i$ from 1 to 2, and decreases by one if recombination changes $j$ from 2 to 1. $z_{ij}^2$ increases or decreases by 3 for each $i$ analogously. As $k$ is the reference marker, it is assumed $z_{ij} = z_{ik}$ before recombination. Therefore $z_{ij} z_{ik}$ decreases by 2 if recombination changes $j$ from 2 to 1, and increases by 1 if recombination changes $j$ from 1 to 2.

This means that, given the recombinant counts $m_{21}$ and $m_{12}$,
\begin{itemize}
\item $n \bar{z}_{j}^*$ increases by $m_{21} - m_{12}$,
\item $s_{jj}^*$ increases by $3 ( m_{21} - m_{12} )$, and
\item $s_{jk}^*$ increases by $m_{21} - 2 m_{12}$.
\end{itemize}
The numerator of Equation \ref{eq:refCorr} consequently becomes
\begin{eqnarray}
  & \left ( s^*_{jk} + m_{21} - 2 m_{12} \right ) - \left( n \bar{z}_{j}^* + m_{21} - m_{12} \right ) \bar{z}_{k}^* \nonumber\\
  & = \left ( s^*_{jk} - n \bar{z}_{j}^* \bar{z}_{k}^* \right ) + m_{21}(1 - \bar{z}_k^*) - m_{12} (2 - \bar{z}_{k}^*) \nonumber\\
  & = \widehat{\sigma}_{jk}^2 + m_{21} (1 - \bar{z}_k^*) - m_{12}( 2 -  \bar{z}_{k}^*) \nonumber \\
  & = \widehat{\sigma}_{jk}^2 + \delta_m (1 - \bar{z}_k^* ) - m_{12}     
\end{eqnarray}
while the variance at marker $j$ changes to
\begin{eqnarray}
  & \left ( s_{jj}^* + 3\delta_m \right ) -  n \left ( \bar{z}_{j}^* + \frac{\delta_m}{n} \right )^2 \nonumber \\
  & = \left ( s_{jj}^* - n (\bar{z}_{j}^*)^2 \right ) + 3 \delta_m - 2 \bar{z}_{j}^* \delta_m -  \frac{\delta_m^2}{n} \nonumber \\
  & = \widehat{\sigma}_{jj}^2 + \delta_m \left ( 3 - 2 \bar{z}_j^* - \frac{\delta_m}{n} \right )
\end{eqnarray}
where $\delta_m = m_{21} - m_{12}$. Equation \ref{eq:refCorr} therefore becomes
\begin{equation}
  \frac{\widehat{\sigma}_{jk}^2 + \delta_m (1 - \bar{z}_k^* ) - m_{12}}{\sqrt{ \widehat{\sigma}_{kk}^2 \Big [ \widehat{\sigma}_{jj}^2 + \delta_m \left ( 3 - 2 \bar{z}_j^* - \frac{\delta_m}{n} \right ) \Big ]}}.
\end{equation}
Noting that $\bar{z}_j^* = \bar{z}_k^* := \bar{z}_{\text{ref}}$ and $s_{jj}^* = s_{kk}^* = s_{jk}^* := s_{\text{ref}}$ by construction, this can be further simplified to
\begin{equation} \label{eq:crossedCorr}
  \frac{ \sigma_{jk}^2 + \delta_m \left(1 - \bar{z}_{\text{ref}}^2 \right ) - m_{12}}{\sqrt{ \widehat{\sigma}_{kk}^2 \Big [ \widehat{\sigma}_{jj}^2 + \delta_m \left ( 3 - 2 \bar{z}_{\text{ref}} - \frac{\delta_m}{n} \right ) \Big ] }}. \\
\end{equation}
The correlation is thereby shown to be changed primarily by the number of recombinant individuals, but that the extent of this change is dictated by the mean additive score of the reference marker, $\bar{z}_{\text{ref}} = \bar{z}_k^*$. Inherited cross overs will always decrease the correlation, though the magnitude of this decrease will depend on $\bar{z}_k^*$ in a non-linear fashion.

\begin{figure}[htp]
  \begin{center}
    \begin{tabular}{cc}
      \includegraphics[scale = 0.6]{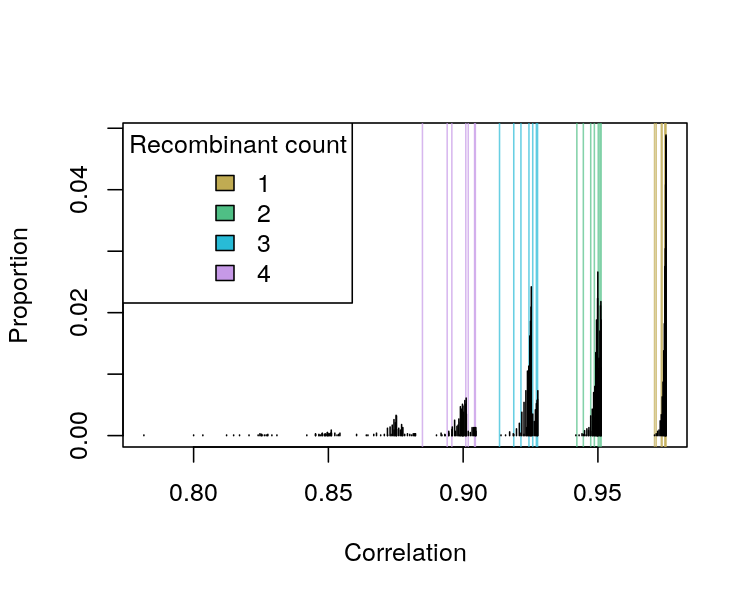} & 
      \includegraphics[scale = 0.5]{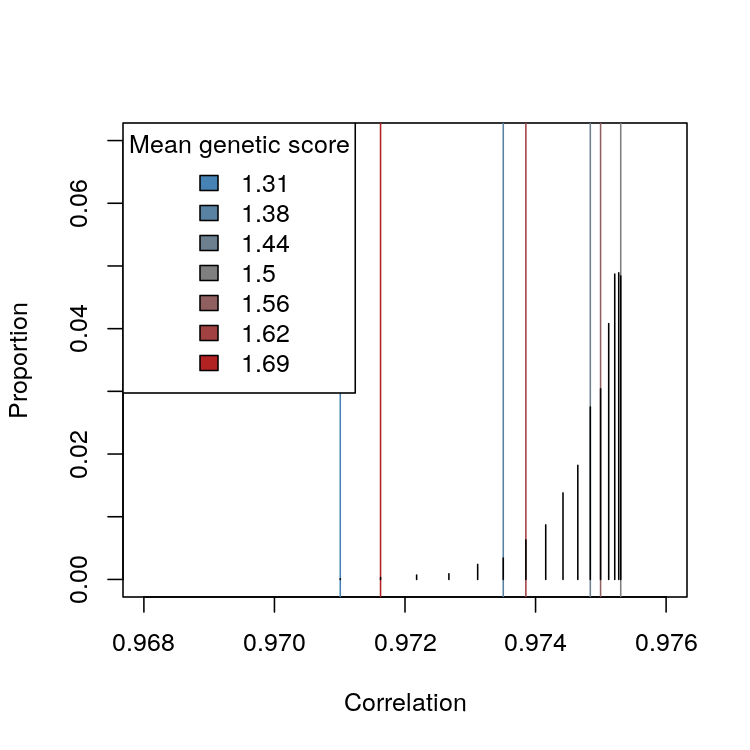} \\
      {\footnotesize (a) Barplot of simulated correlations coloured} & {\footnotesize (b) One recombinant individual} \\
      {\footnotesize by number of recombinant individuals.} & {\footnotesize coloured by mean additive score} \\
      {\footnotesize 15.6\% of populations had a correlation of one.} & \\

    \end{tabular}
  \end{center}
  \caption{The distribution of the correlation between markers Nppa and D4Mit14 over 10,000 simulated BSB crosses.}
  \label{fig:bardetail}
\end{figure}

Using the above, the correlations were computed for the cases of zero, one, two, three, and four recombinant individuals in a population of 80 for a selection of seven mean additive scores. Figure \ref{fig:bardetail} displays these settings with coloured lines laid under the bar plot of Figure \ref{fig:discrete}(b).
These theoretical lines match the values observed in the simulated distribution exactly. As can be deduced from the form of Equation \ref{eq:crossedCorr}, the correlation is dominantly impacted by the number of recombinant individuals, the mean additive score of the reference marker changes the correlation by small amounts in comparison.

\section{Conclusion} \label{sec:conclusion}

This work presents a structural model of genetic measurement incorporating the modern understanding of the genome and typical practice in GWAS. Using this model, the Haldane map distance is shown to be a direct consequence of the structure of the genome and mechanics of inheritance. The model also supports a derivation of genetic correlation, or linkage disequilbrium, which facilitates a comparison of the model's predictions to observed data from \cite{roweetal1994jaxbsb} and \cite{welchetal1996uclabsb}. These comparisons suggest the model fits well for most of the markers examined, and indicate chromosome 4 may have more genetic interference than others in the mouse genome.

Some novel plot matrices were created to support this investigation. The correlation distribution and correlation test plot provide an enriched version of standard correlation matrices by displaying distributional information as well as point estimates. These plots do not scale as well as the standard correlation matrices, but provide an excellent view of the pairwise relationships for a moderate number of markers.

All of these results demonstrate the extraordinary explanatory power and clarity of the model of Section \ref{sec:theModel}. By separating the steps used to create useful genetic data in Figure \ref{fig:modelDiagram}, a rich framework is created. Each step contextualizes the next, tracing a clear path from a structural representation of the genome to the numeric values used in practice. Separated in this way, different methods might be considered for each step.

Selection and annotation are currently based on SNP microarrays as outlined in \cite{laframboise2009}, but as modern genome sequencing advances further this may change, see \cite{heatherchain2016sequencers, hasinetal2017multi, uffelmannetal2021gwas}. The problem of selection will remain important even if a full sequence can be obtained. Minimizing spurious associations would still require the careful choice of regions to compare, which will certainly be annotated for convenience and clarity.

The remaining steps, which involve encoding and summarizing these annotated segments, may be skipped entirely. A plethora of categorical measures of association are outlined in literature such as \cite{goodmankruskal1979measures}. Any of these might be directly applied to annotated genetic data without the need for encoding or summarization.

Finally, this work took a rather narrow view of potential encodings, summaries, and population settings. The model outlined here might immediately be applied to maps such as the dominance map on encoded markers, or to some of the known population settings outlined in Section \ref{sec:correlation} other than the backcross. Further generalization may be possible by treating $\m{F}_X$ and $\m{M}_X$ as random matrices with a known distribution rather than known constants, extending the results gained to a natural population rather than a specified cross. Such extensions may also prove insightful for the sample correlation as outlined in Section \ref{subsec:corrCountCO}.




\bibliographystyle{plain}
\renewcommand*{\bibname}{References} 
\bibliography{../Bibliography/fullbib}

\end{document}